\documentclass[aps,prd,floatfix,superscriptaddress,preprintnumbers,nofootinbib,twocolumn,longbibliography]{revtex4-2}
\pdfoutput=1

\usepackage{amsmath, amssymb, amsfonts, graphicx, mathrsfs, verbatim, float, slashed, bbm, color, xcolor, xspace, enumerate, url, bbding, multirow, outlines,upgreek}
\usepackage[english]{babel}
\usepackage{comment}
\usepackage{tikz,hyperref}
\hypersetup{
     colorlinks   = true,
     citecolor    = red,
     urlcolor     = red,
     linkcolor    = red
}

\definecolor{lime}{HTML}{A6CE39}
\DeclareRobustCommand{\orcidicon}{%
	\begin{tikzpicture}
	\draw[lime, fill=lime] (0,0) 
	circle [radius=0.16] 
	node[white] {{\fontfamily{qag}\selectfont \tiny ID}};
	\draw[white, fill=white] (-0.0625,0.095) 
	circle [radius=0.007];
	\end{tikzpicture}
	\hspace{-3mm}
}

\foreach \x in {A, ..., Z}{%
	\expandafter\xdef\csname orcid\x\endcsname{\noexpand\href{https://orcid.org/\csname orcidauthor\x\endcsname}{\noexpand\orcidicon}}
}

\graphicspath{{figures/}}

\begin{document}

\preprint{SLAC-PUB-17767}
\title{Neutrino and Gamma-Ray Signatures of Inelastic Dark Matter \\ Annihilating outside Neutron Stars}
\author{\orcidA{}~Javier F. Acevedo}
\thanks{\href{mailto:jfacev@stanford.edu}{jfacev@slac.stanford.edu}}
\affiliation{Particle Theory Group, SLAC National Accelerator Laboratory, Stanford, CA 94035, USA}
 \affiliation{Department of Physics, Engineering Physics, and Astronomy, Queen's University, Kingston, Ontario, K7L 2S8, Canada}
 \affiliation{Arthur B. McDonald Canadian Astroparticle Physics Research Institute, Kingston ON K7L 3N6, Canada}
\author{\orcidB{}~Joseph Bramante}
\thanks{\href{mailto:joseph.bramante@queensu.ca}{joseph.bramante@queensu.ca}}
 \affiliation{Department of Physics, Engineering Physics, and Astronomy, Queen's University, Kingston, Ontario, K7L 2S8, Canada}
 \affiliation{Arthur B. McDonald Canadian Astroparticle Physics Research Institute, Kingston ON K7L 3N6, Canada}
\affiliation{Perimeter Institute for Theoretical Physics, Waterloo, Ontario, N2L 2Y5, Canada}
\author{\orcidC{}~Qinrui Liu}
\thanks{\href{mailto:qinrui.liu@queensu.ca}{qinrui.liu@queensu.ca}}
 \affiliation{Department of Physics, Engineering Physics, and Astronomy, Queen's University, Kingston, Ontario, K7L 2S8, Canada}
 \affiliation{Arthur B. McDonald Canadian Astroparticle Physics Research Institute, Kingston ON K7L 3N6, Canada}
\affiliation{Perimeter Institute for Theoretical Physics, Waterloo, Ontario, N2L 2Y5, Canada}
\author{\orcidD{}~Narayani Tyagi}
\thanks{\href{mailto:narayani.tyagi@queensu.ca}{narayani.tyagi@queensu.ca}}
 \affiliation{Department of Physics, Engineering Physics, and Astronomy, Queen's University, Kingston, Ontario, K7L 2S8, Canada}
 \affiliation{Arthur B. McDonald Canadian Astroparticle Physics Research Institute, Kingston ON K7L 3N6, Canada}
\begin{abstract}
We present a new inelastic dark matter search: neutron stars in dark matter-rich environments capture inelastic dark matter which, for interstate mass splittings between about $45 - 285 \ \rm MeV$, will annihilate away before becoming fully trapped inside the object. This means a sizable fraction of the dark matter particles can annihilate while being \emph{outside} the neutron star, producing neutron star-focused gamma-rays and neutrinos. We analyze this effect for the first time and target the neutron star population in the Galactic Center, where the large dark matter and neutron star content makes this signal most significant. Depending on the assumed neutron star and dark matter distributions, we set constraints on the dark matter-nucleon inelastic cross-section using existing H.E.S.S. observations. We also forecast the sensitivity of upcoming gamma-ray and neutrino telescopes to this signal, which can reach inelastic cross-sections as low as $\sim 2 \times 10^{-47} \ \rm cm^2$.
\end{abstract}

\maketitle

\section{Introduction}
The detection of dark matter through its effects on compact astrophysical objects has been investigated in many prior works, see $e.g.$ \cite{Bramante:2023djs} for a review. These have established neutron stars as all-purpose next-generation detectors for any dark matter that interacts with nucleons, since these objects are extremely dense, reach very low temperatures as they age, and accelerate dark matter to semi-relativistic velocities, which altogether offers the possibility of detecting a kinetic heating signature that is robust across a wide range of models
\cite{Baryakhtar:2017dbj,NSvIR:Raj:DKHNSOps,Acevedo:2019agu,NSvIR:SelfIntDM,Bell:2018pkk,NSvIR:GaraniGenoliniHambye,NSvIR:Queiroz:Spectroscopy,NSvIR:Hamaguchi:Rotochemical,NSvIR:Marfatia:DarkBaryon,NSvIR:Bell:Improved,NSvIR:DasguptaGuptaRay:LightMed,NSvIR:GaraniGuptaRaj:Thermalizn,NSvIR:Queiroz:BosonDM,NSvIR:Bell2020improved,NSvIR:zeng2021PNGBDM,NSvIR:anzuiniBell2021improved,NSvIR:Bell2019:Leptophilic,NSvIR:GaraniHeeck:Muophilic,NSvIR:Riverside:LeptophilicShort,NSvIR:Riverside:LeptophilicLong,NSvIR:Bell:ImprovedLepton,NSvIR:Bramante:2021dyx,NSvIR:Fujiwara:2022uiq,NsvIR:Hamaguchi:2022wpz,snowmass:Berti:2022rwn,Alvarez:2023fjj,Bell:2023ysh}. The usage of neutron stars as dark matter detectors has also drawn an increasing interest over the years in the context of asymmetric dark matter that accumulates inside celestial objects and eventually converts them to black holes \cite{Goldman:1989nd,Bertone:2007ae,Kouvaris:2007ay,Kouvaris:2010jy, Kouvaris:2011fi, McDermott:2011jp, Bramante:2013hn, Petraki:2013wwa, Bramante:2013nma, Bertoni:2013bsa, Bramante:2014zca, Autzen:2014tza, Bramante:2015cua, Bramante:2015dfa, Bramante:2016mzo, Bramante:2017ulk,Garani:2018kkd, Acevedo:2019gre, Dasgupta:2020mqg, Acevedo:2020gro, Garani:2021gvc}.
\begin{figure}[t!]
    \centering
    \includegraphics[width=0.95\linewidth]{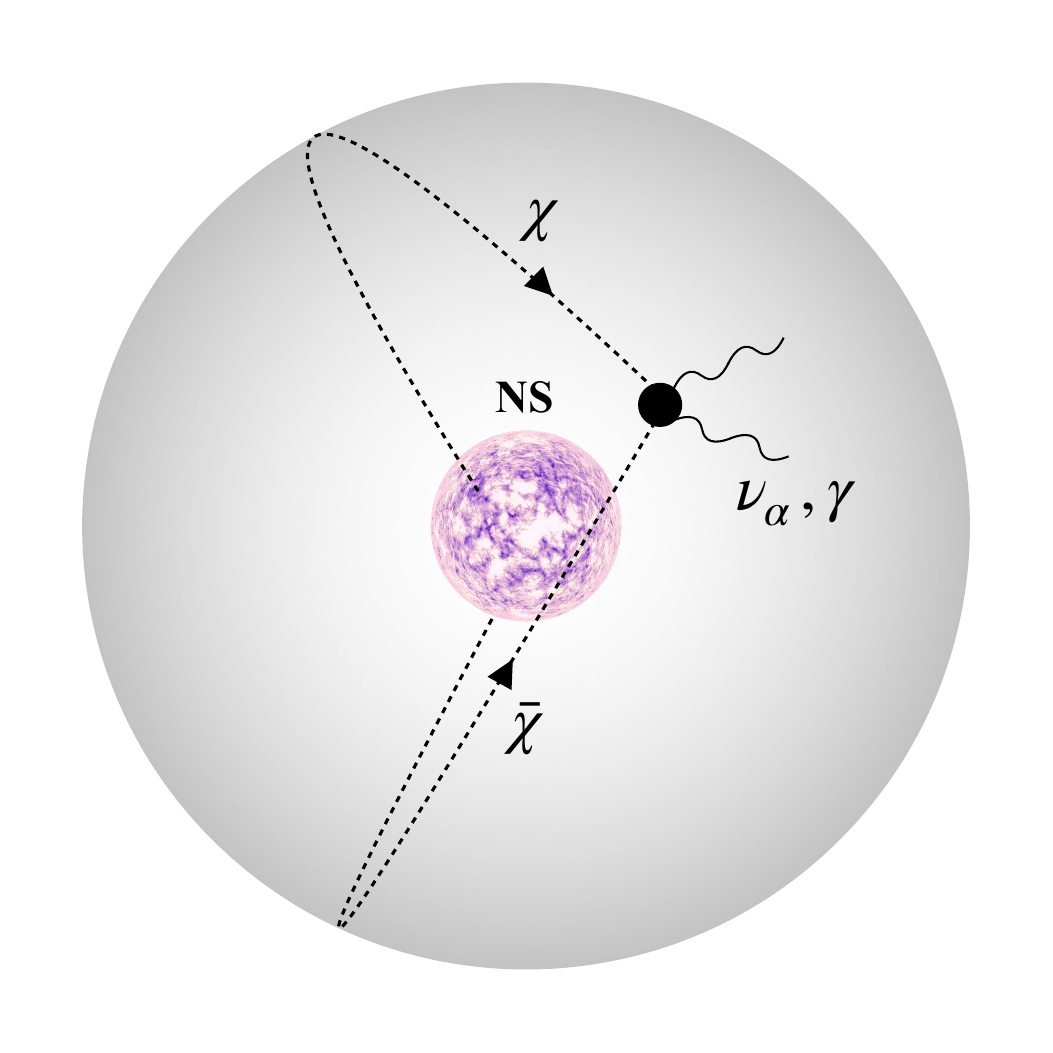}
    \caption{Schematic representation of our effect: captured dark matter that scatters through leading-order inelastic interactions is unable to rapidly thermalize with the neutron star if the inelastic transition becomes forbidden, forming a ``halo" around the object. If the dark matter density at the neutron star position is sufficiently high, a sizable fraction of these particles will annihilate outside before they fully thermalize through loop-level elastic scatterings. This sources a directly observable gamma-ray or neutrino signal.}
    \label{fig:schem}
\end{figure}

Neutron stars also exhibit a unique sensitivity to inelastic dark matter models \cite{Baryakhtar:2017dbj,Bell:2018pkk,Alvarez:2023fjj}, often defined as those with leading-order endothermic interactions of the form $\chi_1 +n \rightarrow \chi_2+n$, where $n$ is the nucleon and $\chi_{1,2}$ are the initial and final dark matter states, with different masses. This scattering process thus requires a threshold energy roughly determined by the mass splitting between both dark matter states \cite{Tucker-Smith:2001myb,Cui:2009xq,Batell:2009vb,Bramante:2016rdh}. Since dark matter is accelerated to a semi-relativistic velocity as it falls into a neutron star, this kinematic condition is met for mass splittings around $\sim 100 \ \rm MeV$, which exceeds what is accessible by direct detection experiments on Earth, of order $\sim 100 \ \rm keV$ \cite{Bramante:2016rdh}. In other words, owing to their strong gravitational field, neutron stars have the potential to probe parameter space for inelastic dark matter that, at present, cannot be reached by other means. \\

In a related but distinct context, it has recently been demonstrated that neutron stars and white dwarfs residing in the Galactic Center can give rise to signals of dark matter annihilating from within, for models with long-lived mediators that escape the object and decay to detectable gamma-rays \cite{Leane:2021ihh,Acevedo:2023xnu}. Due to the high dark matter and stellar density in this region, this ``celestial body focused" \cite{Leane:2021ihh} annihilation is powerful enough to draw constraints on dark matter interactions across a wide mass range, based on existing gamma-ray observations. Subsequently, the analysis of Ref.~\cite{Leane:2021ihh} was extended to the case where the mediators decay to high-energy neutrinos \cite{Bose:2021yhz,Nguyen:2022zwb}, and models for displaced dark matter decays in neutron stars have been explored in \cite{Nguyen:2022zwb,Linden:2024uph}.

In this work, we introduce a new search for dark matter with leading-order inelastic interactions: through its annihilation to high-energy photons and neutrinos \emph{outside} neutron stars. After becoming gravitationally bound to the neutron star, dark matter particles will follow closed orbits through the object, progressively losing energy with each neutron star crossing. Thus, the dark matter eventually loses energy to the point that the inelastic transition becomes kinematically forbidden. For mass splittings approximately in the range $45-285 \ \rm MeV$, this will occur while the orbits still extend beyond the neutron star. Provided that elastic interactions are absent or suppressed, and the background dark matter density is high enough, at any given time there will be an appreciable fraction of dark matter particles annihilating while outside the neutron star volume. We analyze how inelastic dark matter annihilation proceeds in neutron stars in a largely model-independent way, and then provide both constraints and discovery prospects for inelastic dark matter using neutron stars in the Galactic Center as our target. Unlike prior works, this scenario does not require the assumption of long-lived mediators as an intermediate state in the annihilation process. More importantly, this new search may be an alternative to inelastic dark matter searches based on kinetic heating of neutron stars, which at present demand significant observation time with infrared telescopes, are contingent upon the existence of faint radio pulsars close to Earth, and may have substantial background heating sources, which are currently being investigated \cite{Fujiwara:2023hlj,Raj:2024kjq}.

The remainder of this paper proceeds as follows: we begin in Section~\ref{sec:gc_mod} with a discussion of our assumptions concerning the distribution of dark matter and neutron stars in the Galactic Center, our main target for this search, along with the structural properties of a neutron star. In Section~\ref{sec:dmcap}, we briefly review the process of dark matter capture in neutron stars. The partial thermalization of inelastic dark matter with neutron stars is analyzed in Section~\ref{sec:instherm}, including an estimate for the timescale of this process, a discussion of the capture-annihilation equilibrium regime, and the estimated fraction of dark matter that annihilates outside of the neutron star. In Section~\ref{sec:nuandgammasignals}, we outline current and projected sensitivities for finding heavy inelastic dark matter annihilating to gamma-rays and neutrinos outside neutron stars in the Galactic Center. In Section~\ref{sec:dpmodel}, we discuss the class of inelastic dark matter models for which this new search is most effective. Finally, we conclude in Section~\ref{sec:conc}. \\

\section{Galactic Center Modeling}
\label{sec:gc_mod}
To calculate our signal, we require the distribution of both dark matter and neutron stars in the Galactic Center. These inputs determine the overall dark matter annihilation rate proceeding in these objects. Additionally, we also require details about the structure of neutron stars in this region, such as their typical compactness and density profile. These properties ultimately determine the fraction of the total dark matter that will annihilate outside and contribute to the signal.

\subsection{Dark Matter Density and Velocity}
\label{sec:dm_halo}
The dark matter halo profiles we consider are the Navarro-Frenk-White (NFW)~\cite{Navarro:1995iw,Navarro:1996gj}, along with the Einasto~\cite{Graham:2005xx}, and Burkert~\cite{Burkert:1995yz} profiles. These are respectively given by, 
\begin{equation}
  \rho^{\rm NFW}_\chi(R) = \frac{\rho_\chi^0}{\left(\frac{R}{R_s}\right)^\gamma\left(1+\left(\frac{R}{R_s}\right)\right)^{3-\gamma}}~,
\end{equation}
\begin{equation}
  \rho^{\rm Ein}_\chi(R) = \rho_\chi^0 \, \exp{\left[-\frac{2}{\alpha}\left(\left(\frac{R}{R_s}\right)^{\alpha}-1\right)\right]}~, 
\end{equation}
\begin{equation}
   \rho^{{\rm Bur}}_\chi(R) = \frac{\rho_\chi^0}{\left(1+\frac{R}{R_s}\right)\left[1+\left(\frac{R}{R_s}\right)^2\right]}~,
\end{equation}
where $R$ is the Galactocentric distance and $\rho_\chi^0$ is a normalization constant. The parameters were chosen as follows: for the NFW profile, we select a scale radius $R_s = 20 \ \rm kpc$, and consider two different slopes $\gamma =1$ and $\gamma = 1.5$ to approximately span the existing uncertainty on this parameter \cite{necib,Pato:2015dua}. For the latter slope value, we refer to this as a generalized NFW (gNFW) profile. For the Einasto profile, we fix the scale radius $R_s = 20 \ \rm kpc$ and a exponent $\alpha = 0.17$ \cite{Pieri:2009je}. Finally, for the Burkert profile, we choose a scale radius $R_s = 6 \ \rm kpc$, as suggested by some studies \cite{Deason:2012wm}.
In all cases, the normalization factor is determined by requiring the distribution to reproduce the observed dark matter density at the Sun's position, $\rho_\chi^{\odot} \simeq 0.42 \ \rm GeV \ cm^{-3}$ \cite{necib,eilers2019}. 

The dark matter velocity dispersion towards the Galactic Center is not well understood and, in our region of interest, existing uncertainties span a range $v_d \sim 100 - 900 \ \rm km \ s^{-1}$ \cite{sofue:2020rnl}. However, within this range our computed dark matter capture rate only varies by about a factor $\sim 2$. As we show below, this quantity determines the signal strength and, since it is rather insensitive to this input, we assume a fixed velocity dispersion of $v_d \simeq 270 \ \rm km \ s^{-1}$ for simplicity. 

\subsection{Neutron Star Density}
We consider three separate neutron star distribution models presented in Refs.~\cite{Hopman:2006xn,SartNS,Generozov:2018niv}. Ref.~\cite{Hopman:2006xn} put forth a semi-analytic treatment of mass segregation effects around supermassive black holes, and obtained power-law distributions for various stellar object types at the Galactic Center, in the range $10^{-2} - 1 \ \rm pc$. On the other hand, Ref.~\cite{SartNS} conducted a detailed Monte-Carlo simulation study on the Milky Way neutron star population, spanning different assumptions on the Galactic potential and star formation history. In such a work, ten different models were fitted as 4th order polynomials. We take the coefficients given in Table A.1 of this reference, with which the surface densities can be computed. The last one is the ``Fiducial $\times 10$" model at 10~Gyr extracted from Ref.~\cite{Generozov:2018niv}, which used a numerical Fokker-Planck approach combined with existing data on the stellar population in the Galactic Center.

\subsection{Neutron Star Structure}
The strength of the dark matter annihilation signal will also depend on the final size of the dark matter orbits once inelastic scattering becomes kinematically forbidden. To determine these final orbits, we require the neutron star internal structure, which we obtain through the Tolman-Volkoff-Oppenheimer (TOV) equation of hydrostatic equilibrium, combined with an appropriate equation of state. Some of the numerical details of this procedure are discussed in App.~\ref{app:ns_struc}. We use the Brussels-Montreal family of equations of state \cite{Chamel:2010rw,Goriely:2010bm,Potekhin:2013qqa}; specifically its BsK-21 iteration, for which analytical representations were constructed in Ref.~\cite{Potekhin:2013qqa} and are applicable to all neutron star layers. This equation of state indicates agreement with radii and tidal deformability measurements from GW170817 \cite{LIGOScientific:2018cki,Most:2018hfd}, mass and radius fits performed from both gravitational wave and low-energy nuclear data \cite{Raithel:2018ncd}, and X-ray observations of pulsars \cite{Miller:2019cac,Miller:2021qha}. 

We will mostly consider a single benchmark neutron star with mass and radius
\begin{align}
    & M_{\rm NS} \simeq 1.5 M_{\odot}~, \nonumber \\
    & R_{\rm NS} \simeq 12.55 \ \rm km~,
    \label{eq:ns_bench}
\end{align}
as determined by our choice of equation of state. The majority of neutron stars in the Galactic Center are predicted to have masses around this value by the most recent star formation history analysis of this region \cite{2023ApJ...944...79C}. Furthermore, this value is also in agreement with earlier theoretical works \cite{Hopman:2006xn,Alexander:2008tq} and numerical simulations \cite{Zhang:2023cip}. Most neutron stars in this region are expected to have formed in an early star formation episode around $\sim 5 - 10 \ \rm Gyrs$ ago \cite{2020A&A...641A.102S,2023ApJ...944...79C}, and so we will consider their age to be order multi-gigayear. To simplify our analysis, we will also neglect neutron star rotation in our calculations. For a gigayear-old neutron star evolving in isolation, simple estimates using magnetic dipole models already indicate typical periods of order $0.5 - 2 \ \rm s$ \cite{Ridley:2010vr,Bates:2013uma}, which exceeds the dynamical crossing time of captured dark matter. Although this neglects the possibility of pulsar recycling (see $e.g.$ Ref.~\cite{Lorimer:2008se}), we note that the existence of a millisecond pulsar population in this region is currently subject to a high uncertainty. For instance, if a substantial population of these objects were present, it would imply a notable increase in low-mass X-ray binaries compared to observed numbers \cite{Haggard:2017lyq}. In any case, including the effects of rotation would increase the range of inelastic mass splittings accessible to our search. This is because rotation generally allows for more compact neutron star configurations, and therefore greater escape velocities, as the centrifugal force can support heavier masses for a given fixed radius. 

\section{Dark Matter Capture in Neutron Stars}
\label{sec:dmcap}
The gravitational infall and capture of dark matter in neutron stars has been extensively studied in the past \cite{Goldman:1989nd,Kouvaris:2007ay,Bertone:2007ae,Bramante:2016rdh,Bramante:2017xlb,Acevedo:2019agu,Leane:2023woh}. We limit ourselves to providing a brief overview of the process. A dark matter particle is captured when, while traveling through the neutron star, loses energy equal to or greater than the halo kinetic energy it had far from the neutron star. For dark matter in the sub-PeV mass regime, neutron stars are particularly efficient at capture: because of their large escape velocity, a single scatter is sufficient for a dark matter particle to lose enough energy and become gravitationally bound. Above the PeV scale, the initial kinetic energy of the dark matter particles is sufficiently large that multiple scatters are required in a single transit for capture. 

We will focus on the case where the relic dark matter is initially in its lightest state, and scatters with nucleons into a heavier state through a process of the form \cite{Tucker-Smith:2001myb,Cui:2009xq,Batell:2009vb,Bramante:2016rdh}
\begin{align}
    \chi_1 + n \rightarrow \chi_2 + n~,
\end{align}
with a cross-section $\sigma_{\chi n}^{\rm inel}$. Above, $n$ is the scattered nucleon, and $\chi_1$ ($\chi_2$) denotes the lighter (heavier) dark matter state. In Sec.~\ref{sec:dpmodel}, we briefly discuss the class of models for which this scenario is applicable. The interstate mass splitting is defined as
\begin{equation}
    \delta = m_{\chi_2} - m_{\chi_1}~.
\end{equation}
Once excited into the heavy state, the dark matter may either decay back into $\chi_1$ after some time, or undergo exothermic scattering of the form $\chi_2 + n \rightarrow \chi_1 + n$, with an energy loss and cross-section similar to the endothermic case \cite{Batell:2009vb}. For simplicity, from here forward we will refer to $m_{\chi_1}$ as $m_\chi$.    

The number of captured dark matter particles per unit time is calculated as \cite{Bramante:2017xlb}
\begin{equation}
    C_\chi = \sum_{N} C_{N}~,
\end{equation}
where each contribution $C_N$ is the capture rate for dark matter after $N$ scatterings, given by
\begin{widetext}
\begin{eqnarray}
    C_N = \pi {R_{\rm NS}^{2}} \, p_{N}(\tau)\sqrt{\frac{6}{\pi}}\frac{n_{\chi}}{v_{d}}\left[(2v_{d}^{2} + 3v_{\rm esc}^{2}) - (2v_{d}^{2}+3v_{N}^{2})\exp{\left(-\frac{3(v_{N}^{2}-v_{\rm esc}^{2})}{2v_{d}^{2}}\right)}\right]
    \label{eq:cap_N}~.
\end{eqnarray}
\end{widetext}
Above, $n_\chi = \rho_\chi/m_\chi$ is the dark matter number density at the neutron star position, $v_{d}$ is the dark matter halo velocity dispersion, $v_{\rm esc} = \sqrt{2 G M_{\rm NS}/R_{\rm NS}}$ is the escape velocity at the neutron star surface measured by a local observer, and $v_N$ is the dark matter velocity after it has undergone $N$ scatters. Our assumptions concerning the dark matter density and velocity dispersion in the Galactic Center are discussed in Sec.~\ref{sec:dm_halo}. The dark matter velocity after $N$ scatters is determined by the fractional energy loss in each subsequent scatter. Unless the mass splitting is close to the maximum value allowed by kinematics, the fraction of energy lost in an inelastic scattering will be approximately the same as a the fraction lost in a pure elastic scattering. For instance, for mass splittings up to about $90\%$ of the maximum value accessible to a neutron star, the energy lost in an elastic scattering differs from the inelastic case by less than $10\%$. Thus, to simplify our calculations, we follow the calculations of Ref.~\cite{Bramante:2017xlb} and take $v_N \equiv v_{\rm esc}(1-\beta_{+}/2)^{-N/2}$, where $\beta_{+}=4m_{\chi}m_{n}/(m_{\chi}+m_{n})^{2}$ as determined by elastic scattering kinematics and $m_n \simeq 0.93 \ \rm GeV$ is the nucleon mass. Finally, the factor $p_N (\tau)$ is the probability for capture after $N$ scatters, given by
\begin{equation}
    p_N(\tau) = \frac{2}{N!} \int_{0}^{1} \left(\cos\theta\right)^{N+1} \tau^N \exp\left(-\tau \cos\theta\right)\ d\left(\cos\theta\right)~,
    \label{eq:cap_pN}
\end{equation}
where $\tau = 3 \sigma^{\rm inel}_{\chi n} M_{\rm NS} / 2 \pi R_{\rm NS}^2 m_n$ is the optical depth. This is the ratio between the inelastic dark matter-nucleon cross-section and the cross-section at which dark matter scatters once on average along a distance $2R_{\rm NS}$. 

It is also useful to estimate the so-called ``saturation" cross-section for which almost all the incoming dark matter is captured by a neutron star (strictly speaking, all of the flux is captured only for an infinitely large cross-section). Above this value, the annihilation signal reaches its peak, defining the ultimate sensitivity of our search. For dark matter between GeV and PeV mass, a single scatter is sufficient for capture. Thus, requiring $\sigma^{\rm inel}_{\chi n} n_n R_{\rm NS} \sim 1$, with $n_n \sim 3 M_{\rm NS}/4 \pi m_n R_{\rm NS}^3$, provides a reasonable estimate of this cross-section. Above the PeV scale, the above estimate is modified by the fact that $N$ scatters are needed for a dark matter particle to be captured. In this regime, this cross-section must linearly increase with mass, since the dark matter's halo kinetic energy also increases linearly with mass. For our benchmark neutron star, this cross-section is
\begin{equation}
    \sigma^{\rm sat}_{\chi n} \simeq 3 \times 10^{-45} \ {\rm cm^2} \times \max\left[1,\frac{m_\chi}{\rm PeV}\right]~,
    \label{eq:sigma_sat}
\end{equation}
where the $\max$ function reflects the transition between the single- and the multi-scatter capture regime.  

\section{Inelastic Dark Matter Thermalization}
\label{sec:instherm}
Once a dark matter particle is captured, it will progressively lose energy from repeated scatters as it periodically crosses the neutron star. In the absence of any scattering suppression, energy loss proceeds until the dark matter particle settles inside, with a final energy roughly determined by the neutron star temperature. For inelastic dark matter, however, this process may stall before the particle reaches an orbit fully contained within the neutron star volume, if the kinematic threshold for inelastic scattering stops being met at some intermediate energy. Below, we analyze the range of mass splittings and timescales over which this occurs, as well as the distribution of partially thermalized dark matter.

\subsection{Kinematic Threshold}
In the regime $m_\chi \gg m_n$, Pauli blocking effects are negligible and we may approximate the neutron targets to initially be at rest in a dark matter-nucleon collision. In this limit, the center-of-mass energy is simply
\begin{equation}
    E_{\rm CM} = \left(m_\chi^2 + m_n^2 + \frac{2 m_\chi m_n}{\sqrt{1-v^2}}\right)^{1/2}~,
    \label{eq:inel_thresh}
\end{equation}
where $v$ is the dark matter velocity as measured by a local observer in the neutron star. Neglecting angular momentum, this velocity parameterized in terms of the Schwarzchild radial coordinate $r$ (see App.~\ref{app:dm_orb}) reads
\begin{equation}
    v(r) = \left(1 - \frac{g_{tt}(r)}{g_{tt}(r_f)}\right)^{1/2}~.
    \label{eq:v_loc_main}
\end{equation}
Above $g_{tt}(r)$ is the first diagonal component of the neutron star metric (see App.~\ref{app:ns_struc}), and $r_f$ is the turning point of the dark matter particle. For captured dark matter, $r_f$ is finite since it is gravitationally bound to the object. On the other hand, for incoming dark matter particles from the halo one may take $r_f \rightarrow \infty$ and $g_{tt}(r_f) \rightarrow 1$, which amounts to neglecting the initial halo kinetic energy. Since neutron stars have semi-relativistic escape velocities which are much larger than the halo velocity, this is a reasonable approximation. The relation between $r_f$ and the total energy of the dark matter particle, including the gravitational binding contribution, is specified further below (see also App.~\ref{app:dm_orb}). For the dark matter to endothermically scatter into the heavier state, we require the total center-of-mass energy of the collision to be
\begin{equation}
    E_{\rm CM} \gtrsim m_n + m_\chi + \delta~.
    \label{eq:inel_thresh_2}
\end{equation}
Eq.~\eqref{eq:inel_thresh_2} determines the kinematic threshold for inelastic scattering to proceed.

\subsection{Accessible Mass Splittings}
With the kinematic condition defined, we estimate the mass splitting range for which the dark matter will only partially thermalize through inelastic interactions. To this end, we first analyze the various energy scales that are relevant for the thermalization process. 

In what follows, we define the energy $\varepsilon$ of a dark matter particle as the total energy including gravitational binding contribution, see App.~\ref{app:dm_orb}. Note that $\varepsilon$ is in fact $\leq m_\chi$ when the dark matter particle is captured by the neutron star, with the equality setting the boundary between bound and open orbits. In App.~\ref{app:dm_orb} we show that, for a captured dark matter particle, the turning point $r_f$ is related to this quantity through the relation 
\begin{equation}
    g_{tt}(r_f) = \left(\frac{\varepsilon}{m_\chi}\right)^2~.
    \label{eq:turn_point_out_0}
\end{equation}
If the turning point is outside the neutron star volume, this relation simplifies to
\begin{equation} 
    r_f = \frac{2 G M_{\rm NS}}{1 - (\varepsilon/m_\chi)^2}~.
    \label{eq:turn_point_out_1}
\end{equation}

\begin{figure}[t]
    \centering
    \includegraphics[width=\linewidth]{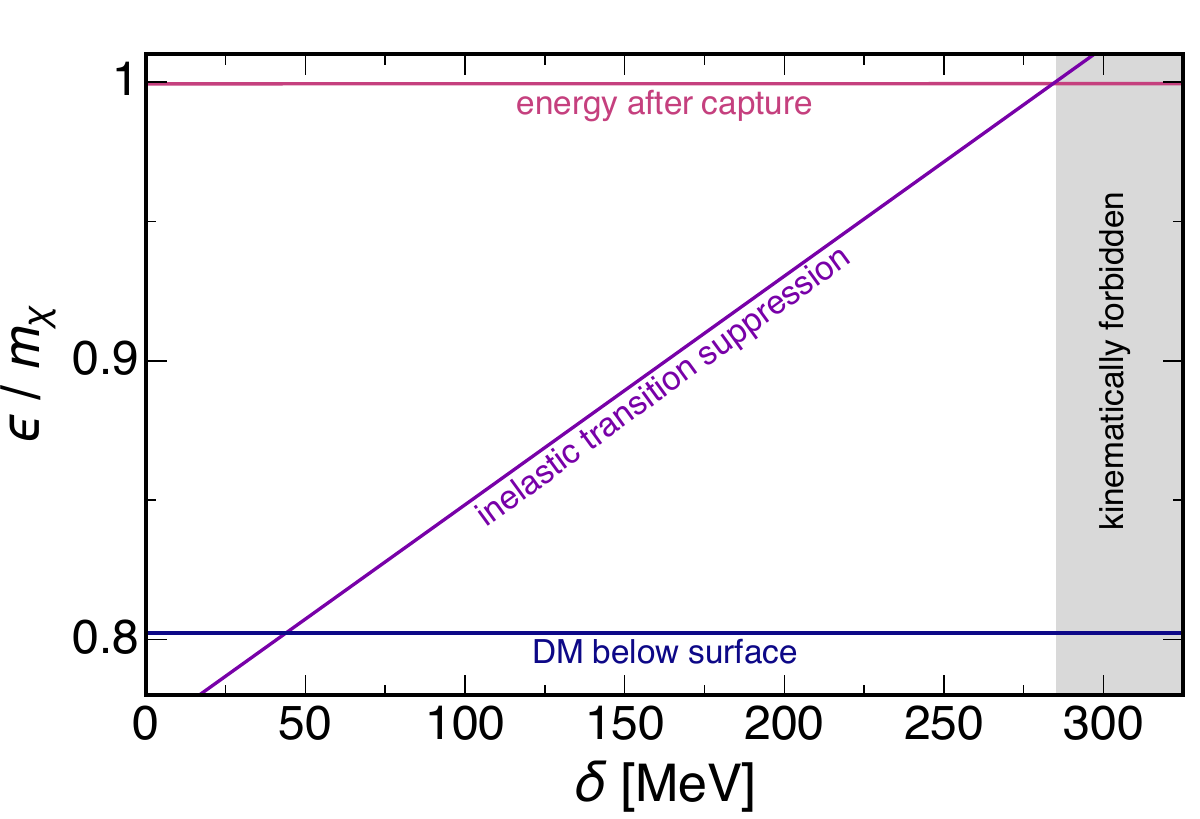}
    \caption{Dark matter energy immediately after capture, once the inelastic transition becomes kinematically inaccessible during thermalization, and once the orbits are fully contained within our benchmark neutron star. We also indicate in the shaded area the interstate splitting range where the capture rate is suppressed due to inelastic scattering being kinematically forbidden.}
    \label{fig:thermenergies}
\end{figure}

In total, there are three energy scales that must be considered: 
\begin{enumerate}
    \item Prior to capture, the dark matter has an energy $\varepsilon \simeq m_\chi + m_\chi v_d^2/2$ far from the neutron star. Upon capture after $N$ inelastic scatters, the energy of the particle drops to $\varepsilon = m_\chi$. In other words, just the kinetic energy the particle had far from the neutron star is lost. We then require at least $(N+1)$ inelastic scatterings to obtain an orbit with a finite turning point, $cf.$ Eq.~\eqref{eq:turn_point_out_1}. The average recoil energy for inelastic scattering in the regime where $\delta \lesssim m_n \ll m_\chi$ is \cite{Bell:2018pkk}
    \begin{equation}
        \langle \Delta E^{\rm inel}_{\rm rec} \rangle \simeq \frac{m_n m_\chi^2 \gamma^2 v^2 \left(1 - (\gamma +1)(\delta/m_n)/\gamma^2 v^2\right)}{m_n^2 + m_\chi^2 + 2 \gamma m_n m_\chi}~,
    \end{equation}
    where $\gamma^{-1} = \sqrt{1 - v^2}$. Thus, we take the energy at the beginning of the thermalization process to be of order
    \begin{equation}
        \varepsilon_i \simeq m_\chi - \langle \Delta E^{\rm inel}_{\rm rec} \rangle - \delta~,
    \end{equation}
    where we evaluate $\langle \Delta E^{\rm inel}_{\rm rec} \rangle$ at its minimum value, which is near the neutron star surface.
    
    \item The captured dark matter particles further lose energy through inelastic scatters until the kinematic threshold is no longer met. To obtain the energy at which this occurs, we first solve for $v$ in Eq.~\eqref{eq:inel_thresh_2} using the center-of-mass energy Eq.~\eqref{eq:inel_thresh}. For simplicity, we provide here the result in the limit $\delta \lesssim m_n \ll m_\chi$,
    \begin{equation}
       \ \ \ \ v_{\rm min} \simeq \sqrt{2}\left(\frac{\delta}{m_n}\right)^{1/2} - \frac{3}{2\sqrt{2}} \left(\frac{\delta}{m_n}\right)^{3/2}~.
    \end{equation}
    This expansion is accurate within $\lesssim 10 \%$ through the mass splitting range accessible to a typical neutron star. Using Eqs.~\eqref{eq:v_loc_main} and \eqref{eq:turn_point_out_0}, we solve for the energy, which yields
    \begin{equation}
        \varepsilon_m \simeq m_\chi \left(\frac{g_{tt}(0)}{1-v_{\rm min}^2}\right)^{1/2}~.
        \label{eq:energ_interm}
    \end{equation}
    Above, we have explicitly taken $r = 0$ in Eq.~\eqref{eq:v_loc_main}, so that the center-of-mass energy is maximized. This estimate indicates the energy at which a dark matter particle even passing through the center, where the highest velocity is attained, would not meet the threshold for inelastic scattering. 
    \item Once inelastic scattering becomes suppressed, if the dark matter is able to elastically scatter at the loop-level, further energy loss proceeds at a much slower rate until the dark matter is eventually fully contained within the neutron star. This occurs when $r_f = R_{\rm NS}$, or equivalently a final energy \begin{equation}
    \varepsilon_{\rm f} = m_{\chi} \left({1-\frac{2 G M_{\rm NS}}{R_{\rm NS}}}\right)^{1/2}~,
    \end{equation}
    $cf.$ Eq.~\eqref{eq:turn_point_out_1}. Numerically, $\varepsilon_f \simeq 0.8 \, m_\chi$ for our benchmark neutron star, implying that only a small fraction of the energy relative to the rest mass is lost before the dark matter becomes fully trapped.  
\end{enumerate}
Figure~\ref{fig:thermenergies} shows each of these energy scales for our benchmark neutron star, $cf.$ Eq.~\eqref{eq:ns_bench}.
The ``inelastic transition suppression" line denotes where Eq.~\eqref{eq:inel_thresh_2} is violated when the dark matter particles are transiting through the center of the neutron star, where they attain the maximum velocity. The two remaining lines indicate the typical energy once a dark matter particle is captured, and the energy once it becomes fully contained within the neutron star. The intersection of these with the inelastic transition suppression line approximately determines the mass splitting range for which we obtain a sizable fraction of inelastic dark matter annihilating outside the neutron star. On the one hand, mass splittings below $\delta_{\rm min} \simeq 45 \ \rm MeV$ result in the dark matter thermalizing through inelastic scatterings to the point that is contained inside the neutron star. On the other hand, for mass splittings above $\delta_{\rm max} \simeq 285 \ \rm MeV$, the energy of the dark matter falling onto the neutron star is insufficient for inelastic scattering to initially proceed. This maximum mass splitting is highlighted by the shaded area. 

For completeness, we briefly discuss how this mass splitting window varies with neutron star mass and radius, always assuming a BsK-21 equation of state. The lightest neutron stars in the Milky Way's inner parsec are expected to be of order $M_{\rm NS} \simeq 1.4 M_\odot$ \cite{2023ApJ...944...79C}. On the other hand, the heaviest non-rotating neutron star predicted by our choice of equation of state has a mass of order $M_{\rm NS} \simeq 2.2M_\odot$ \cite{Potekhin:2013qqa}. Repeating the above analysis, we obtain mass splitting windows $(\delta_{\rm min}, \delta_{\rm max}) \simeq $ $(40,260) \ \rm MeV$ and $(118,672) \ \rm MeV$ for the lightest and heaviest neutron star, respectively. Both $\delta_{\rm min}$ and $\delta_{\rm max}$ increase with the neutron star mass because a stronger gravitational field implies that more center-of-mass energy is available to excite the inelastic transition. Thus, $\delta_{\rm min}$ must increase for the dark matter to remain outside after being captured, while $\delta_{\rm max}$ also increases as it becomes possible to capture dark matter with larger interstate splittings. In practice though, we expect most neutron stars will have masses around our chosen benchmark, $cf.$ Eq.~\eqref{eq:ns_bench}, so we take $(\delta_{\rm min}, \delta_{\rm max}) \simeq (45,285) \ \rm MeV$ as our viable mass splitting range.

\subsection{Thermalization Timescale}
We now consider the timescale for the captured dark matter particles to thermalize from the energy $\varepsilon_m$ at which inelastic scattering stops, to the final energy $\varepsilon_f$ at which they are fully contained within the neutron star. This constitutes a lower bound on the total time required for a particle to fully thermalize starting from the initial energy $\varepsilon_i$ it has immediately after capture. We compute this lower bound using existing thermalization calculations with elastic interactions \cite{Acevedo:2019agu}. A more exact computation including the initial inelastic thermalization phase would demand accounting for the progressive reduction of phase space available for inelastic scattering as the dark matter loses energy, along with a more detailed modeling of individual particle trajectories, which is outside the exploratory scope of this work. In practice, we require dark matter particles not to become trapped within the neutron star in a short timescale to allow for sizable annihilation outside. Further below, we use this lower bound on that timescale to conservatively estimate how suppressed the elastic cross-section must be relative to the inelastic one for this condition to be met.

Following the approach of Ref.~\cite{Acevedo:2019agu}, we estimate the energy loss rate of a captured dark matter particle as 
\begin{equation}
    \frac{d\varepsilon}{dt} = - \frac{\langle \Delta \varepsilon \rangle_{\rm NS}}{t_{\rm orb}(\varepsilon)}~,
    \label{eq:dm_therm_main}
\end{equation}
where $t_{\rm orb}(\varepsilon)$ is the orbital period as a function of energy (see App.~\ref{app:dm_orb}), and we have introduced the energy loss averaged over the neutron star volume as  
\begin{equation}
    \langle \Delta{\varepsilon} \rangle_{\rm NS} \simeq \frac{2}{Z_{\rm NS}} \int_{0}^{Z_{\rm NS}} \, \sigma_{\chi n}^{\rm elas} \, \langle E^{\rm elas}_{\rm rec}(z) \rangle \, n_n(z) \, dz~,
    \label{eq:dE_ave_NS}
\end{equation}
where $dz = \sqrt{-g_{rr}(r)} \, dr$ is the proper depth element of the neutron star, $Z_{\rm NS}$ is the integrated proper radius (valid for a stationary metric), and $\sigma_{\chi n}^{\rm elas}$ is the elastic dark matter-nucleon cross-section, which we assume to be smaller than its inelastic counterpart. The factor $\langle E^{\rm elas}_{\rm rec}(z) \rangle$ is the average energy transfer for elastic scattering, which is a function of depth as this determines the kinetic energy. This is given by \cite{Baryakhtar:2017dbj}
\begin{equation}
   \langle E^{\rm elas}_{\rm rec} \rangle = \frac{m_n m_\chi^2 \gamma(r)^2 v(r)^2}{m_n^2 + m_\chi^2 + 2 \gamma(r) m_n m_\chi}~,  
\end{equation}
where $v(r)$ is given by Eq.~\eqref{eq:v_loc_main}. The additional factor of 2 in Eq.~\eqref{eq:dE_ave_NS} accounts for the dark matter particles transiting all neutron star layers twice.

We then integrate Eq.~\eqref{eq:dm_therm_main} from $\varepsilon = \varepsilon_m$ to $\varepsilon = \varepsilon_f$, which gives the approximate timescale for which dark matter is trapped within the neutron star exclusively through loop-suppressed elastic interactions
\begin{equation}
    t_{\rm therm} \simeq -\int_{\varepsilon_m}^{\varepsilon_f} \frac{t_{\rm orb}(\varepsilon)}{\langle \Delta{\varepsilon} \rangle_{\rm NS}} \, d\varepsilon~.
    \label{eq:therm_bound}
\end{equation}
Below, we compare this timescale to the time it takes for the dark matter to annihilate while its orbit is extends beyond the neutron star volume. We remark that we have neglected here any coherent nuclear enhancement factors that occur in the crust of the neutron star. While these enhancement factors can be significant \cite{Acevedo:2019agu}, the overall optical depth of the crust is usually a subdominant contribution to the total optical depth of the star. 

\subsection{Capture-Annihilation Equilibrium}
The total number of dark matter particles bound to the neutron star will evolve according to 
\begin{equation}
    \frac{dN_\chi}{dt} = C_{\chi}-\frac{\langle \sigma_{\rm ann} v_{\rm rel} \rangle N_\chi^2}{V}~,
    \label{eq:Nchi_dot}
\end{equation}
where $\langle \sigma_{\rm ann} v_{\rm rel} \rangle$ is the thermally averaged dark matter annihilation cross-section $\sigma_{\rm ann}$ multiplied by the relative velocity $v_{\rm rel}$, and $V$ is the proper volume where annihilation proceeds. For a recent discussion on partially thermalized dark matter annihilation with velocity- and momentum transfer-dependent operators see Ref.~\cite{Bell:2023ysh}. For inelastic dark matter that has only been partially thermalized with a neutron star, both the relative velocity and the volume will be approximately determined by the characteristic size of the orbits once the inelastic transition becomes suppressed. The latter is obtained through the integration of $dV = 4 \pi r^2 dr / \sqrt{-g_{rr}(r)}$ over the interval defined by the turning point of the dark matter particles once the inelastic transition becomes suppressed. 

In our analysis, we expand the annihilation cross-section in terms of partial waves as
\begin{equation}
    \langle \sigma_{\rm ann} v_{\rm rel} \rangle = \langle \sigma_{\rm ann} v_{\rm rel}\rangle_0 \, \sum_{l=0}^{\infty} a_l \, v_{\rm rel}^{2l},
    \label{eq:sigma_ann_partial}
\end{equation}
and analyze each contribution individually by setting $a_{l} = 1$ for some specific mode $l$ and setting all remaining coefficients to zero. As our benchmark, we will take $\langle \sigma_{\rm ann} v_{\rm rel}\rangle_0 = 3 \times 10^{-26} \ \rm cm^3 \ s^{-1}$. We will focus on the most commonly considered $s$-wave ($l=0$) and $p$-wave ($l=1$) modes.

For the initial condition $N_\chi(t = 0) = 0$, the solution to Eq.~\eqref{eq:Nchi_dot} is
\begin{equation}
    N_{\chi}(t) = N_{\rm eq} \tanh\left(\frac{t}{t_{\rm eq}}\right)~,
\end{equation}
where we have defined the capture-annihilation equilibrium timescale as
\begin{equation}
    t_{\rm eq} = \left(\frac{V}{C_{\chi} \langle \sigma_{\rm ann} v_{\rm rel} \rangle}\right)^{1/2}~.
    \label{eq:t_eq_main}
\end{equation}
Once $t \gtrsim t_{\rm eq}$, the number of dark matter particles reaches an equilibrium value approximately given by
\begin{equation}
    N_{\rm eq} \simeq C_{\chi} \, t_{\rm eq}~,
    \label{eq:N_eq_main}
\end{equation}
and dark matter annihilation must then proceed at a steady-state rate given by 
\begin{equation}
    \Gamma_{\rm ann} = \frac{C_\chi}{2}~,
\end{equation}
where the factor of $2$ reflects that two dark matter particles are depleted per annihilation event.
\begin{figure}[t]
    \centering
\includegraphics[width=\linewidth]{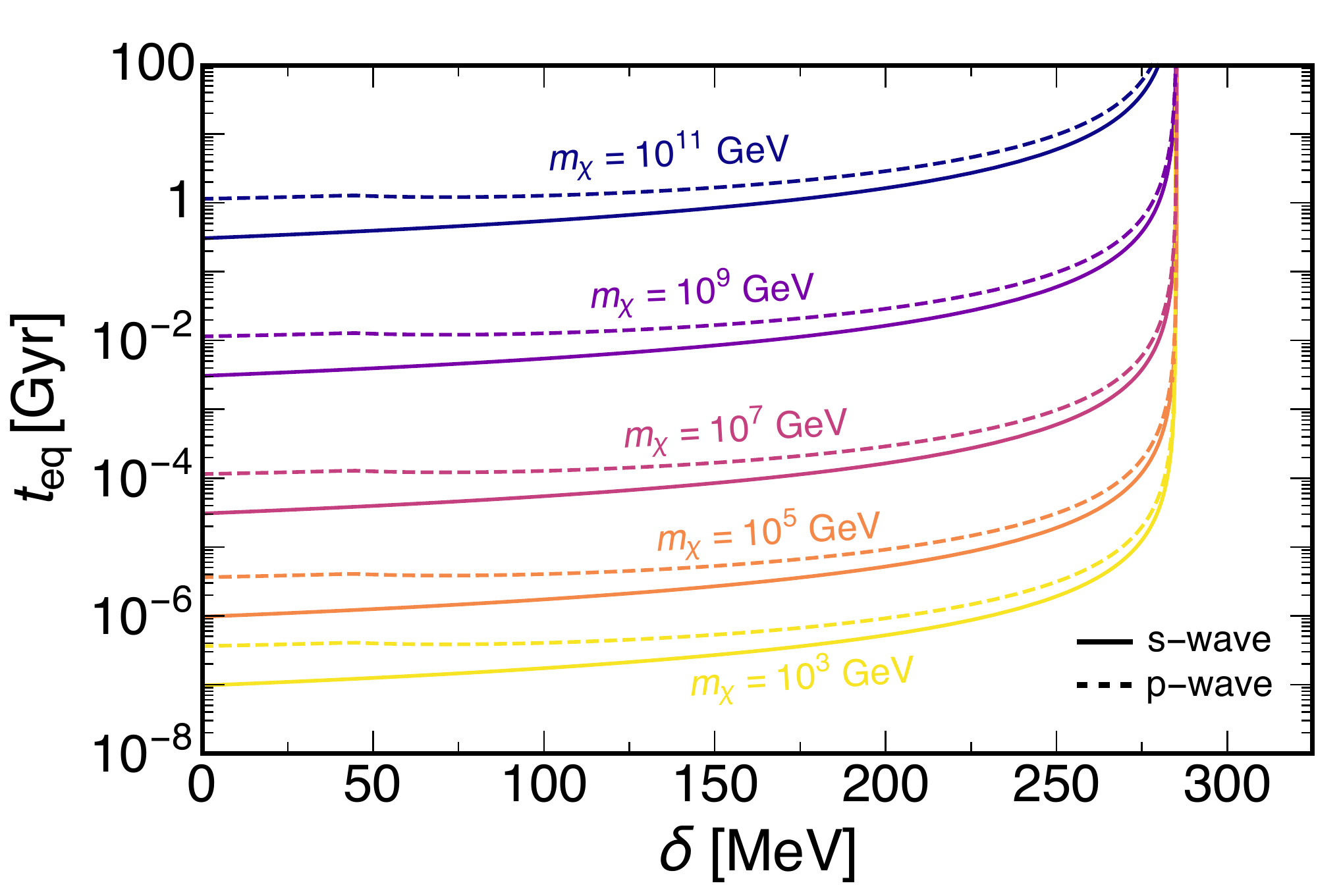}
    \caption{Capture-annihilation equilibrium timescale for our benchmark neutron star and various dark matter masses as indicated, for $s$-wave and $p$-wave channels. For this plot, we have taken a dark matter velocity dispersion $v_{d} = 270 \ \rm km \ s^{-1}$ and density $\rho_\chi \simeq 7.25 \times 10^3 \ \rm GeV \ cm^{-3}$, the latter corresponding to an NFW profile at $R \simeq 1 \ \rm pc$. The inelastic cross-section has been fixed to $\sigma^{\rm inel}_{\chi n} = 10^{-45} \ \rm cm^2$, approximately the lowest value for producing a potentially observable gamma-ray signal with upcoming telescopes under these assumptions.}
    \label{fig:cap-ann-eq}
\end{figure}

Figure~\ref{fig:cap-ann-eq} shows the $s$-wave and $p$-wave equilibration timescale as a function of mass splitting for our benchmark neutron star. In the $p$-wave case, we take the relative velocity as approximately the dark matter velocity averaged over one orbital cycle. For the capture rate, we have taken $\rho_\chi \simeq 7.25 \times 10^3 \ \rm GeV \ cm^{-3}$, corresponding to a standard NFW profile at $R = 1 \ \rm pc$ from the Galactic Center, and an inelastic cross-section $\sigma_{\chi n}^{\rm inel} \simeq 10^{-45} \ \rm cm^2$, corresponding to approximately the lowest value for which a prospective gamma-ray signal might be observed by future gamma-ray telescopes. These values fix the capture rate to about the lowest value for which even neutron stars near the edge of our region of interest would contribute to an observable signal. For these conservative assumptions, we see that capture-annihilation equilibrium is reached within typical neutron star ages for all the mass range analyzed, so long as $\delta$ is not too close to $\delta_{\rm max}$. When $\delta \rightarrow \delta_{\rm max}$, the turning point of the dark matter particle orbit diverges, which is equivalent to no particle being captured. As a result, the equilibration time diverges in this limit. 

Finally, we address the effects of loop-level elastic scattering on the annihilation equilibrium. For a sizable amount of dark matter to annihilate outside the neutron star, the timescale for it to fully sink below the surface must be longer than the timescale it takes to achieve capture-annihilation equilibrium. This hierarchy between timescales is ultimately determined by how suppressed the elastic cross-section is relative to its inelastic counterpart. We provide an estimate of the suppression required by comparing the time to thermalize $t_{\rm therm}$ to $t_{\rm eq}$, where we evaluate the latter at $R = 1 \ \rm pc$ for an NFW profile. This is extremely conservative, since most of the signal is produced by neutron stars much closer to the Galactic Center. Such neutron stars achieve capture-annihilation equilibrium on a timescale much shorter than those at $\sim \ \rm pc$ distances. Moreover, as mentioned above our thermalization timescale is in fact an underestimate, since we do not account for the initial inelastic thermalization stage. Under these conservative assumptions, equating $t_{\rm therm} = t_{\rm eq}$ and solving for the elastic cross-section, we obtain
\begin{align}
    \sigma^{\rm elas}_{\chi n} \lesssim & \, \, \, 10^{-57} \ {\rm cm^2} \, F(\delta)\left(\frac{\rho_{\chi}}{\rho^{\rm NFW}_{\chi}(R = 1 \ {\rm pc})}\right)^{1/2} \\ & \times \left(\frac{\sigma^{\rm inel}_{\chi n}}{10^{-45} \ \rm cm^2}\right)^{1/2} \max\left[\left(\frac{m_\chi}{\rm PeV}\right)^{1/2} \! \! , \, 1\right] \nonumber
    \label{eq:cs_hierar}
\end{align}
for our benchmark neutron star. To derive this, we have used the linear and inverse scaling of $t_{\rm therm}$ ($cf.$ Eq.~\eqref{eq:therm_bound}) with dark matter mass and elastic cross-section, respectively. We have also used the approximate linear scaling of the capture rate with the ratio $\sigma^{\rm inel}_{\chi n}/\sigma^{\rm sat}_{\chi n}$, which enters $t_{\rm eq}$. The function $F(\delta)$ encapsulates the remaining dependence on the mass splitting and must be computed numerically. This function diverges for $\delta \rightarrow \delta_{\rm min, max}$. This is because when $\delta \rightarrow \delta_{\rm min}$, the dark matter is already almost completely thermalized through inelastic scattering. On the other hand, if $\delta \rightarrow \delta_{\rm max}$, the capture-annihilation equilibration time diverges because the turning point of the dark matter particles becomes infinite, making the volume where annihilation proceeds infinitely large. However, for mass splittings $\sim 10\%$ above $\delta_{\rm min}$ and $\sim 90\%$ below $\delta_{\rm max}$, $F(\delta)$ ranges between $10^{-2}$ and $1$. We reiterate that higher elastic cross-section values are allowed without impacting on the signal strength, as Eq.~\eqref{eq:t_eq_main} was evaluated at $R = 1 \ \rm pc$ and we have used a lower bound on the actual thermalization time. In particular, neutron stars closer to the Galactic Center, due to the higher dark matter density, have much shorter capture-annihilation equilibrium times and thus admit much less supressed elastic cross-sections while still having a sizable amount of dark matter annihilating outside. In Sec.~\ref{sec:dpmodel}, we discuss what class of models generate elastic cross-sections that are suppressed relative to their inelastic counterpart. 

\subsection{Annihilation Rate Outside}
With the timescales for thermalization and capture-annihilation equilibrium outlined, we now turn to estimating the fraction of dark matter that will annihilate outside the neutron star volume. As we discuss in App.~\ref{app:inel_prof}, we approximate the number density of dark matter particles seen by a local observer as
\begin{equation}
    n_\chi(r) = n_\chi^{0} \left(1 - v(r)^2\right)^{1/2}~,
    \label{eq:DMdens_main} 
\end{equation}
where $n_\chi^{0}$ is a normalization constant determined by the steady-state number of captured dark matter particles. The orbital velocity $v(r)$ is evaluated for a fixed particle's energy $\varepsilon = \varepsilon_m$ since we are focusing on the regime where $t_{\rm eq} \ll t_{\rm therm}$. In other words, once the inelastic transition can no longer be excited, the dark matter will annihilate long before the orbit changes significantly through suppressed elastic scatterings. The normalization factor is, in fact, irrelevant for determining the fraction of the total annihilation rate occurring outside. Rather, for turning points $r_f \geq R_{\rm NS}$ we are interested in the ratio  
\begin{figure}[t]
    \centering
    \includegraphics[width=\linewidth]{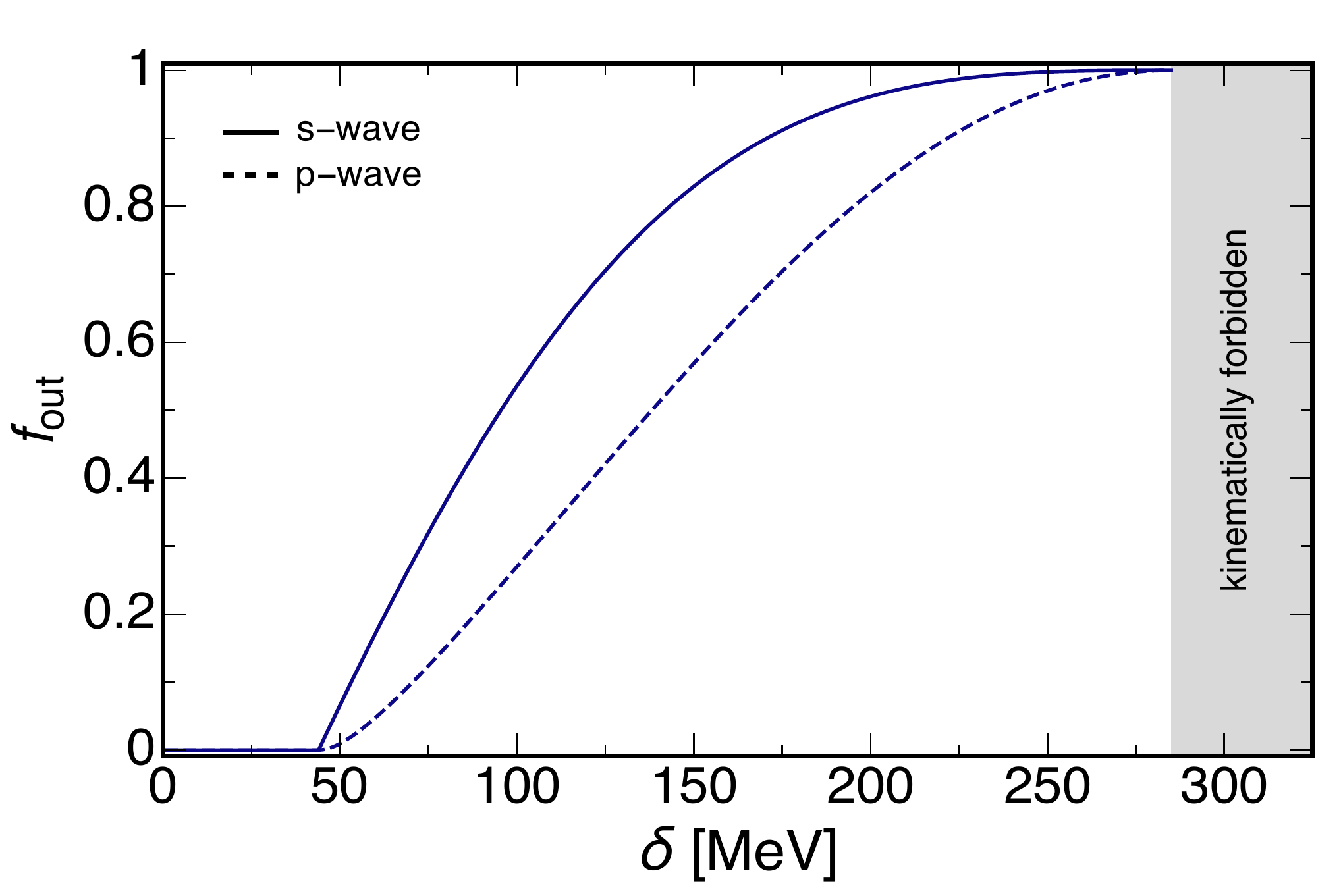}
    \vspace{-5mm}
    \caption{Fraction of the total annihilation rate occurring outside the neutron star volume as a function of interstate splitting, for $s$-wave and $p$-wave annihilation channels as indicated. The shaded region indicates the maximum mass splitting for which the inelastic transition can be excited when the dark matter initially approaches the neutron star at approximately the escape velocity.}
    \label{fig:ann_frac}
\end{figure}

\begin{equation}
 f_{\rm out} = \frac{\int_{R_{\rm NS}}^{r_f}  n_\chi^2(r) \langle \sigma_{\rm ann} v_{\rm rel} \rangle \, dV} {\int_0^{r_f}  n_\chi^2(r) \langle \sigma_{\rm ann} v_{\rm rel} \rangle \, dV } ~,
    \label{eq:annfrac-cond}
\end{equation}
where $f_{\rm out} \in [0,1]$ is the estimated fraction of the annihilation rate that proceeds outside the neutron star volume. As before, we expand $\langle \sigma_{\rm ann} v_{\rm rel} \rangle$ into partial waves and individually analyze each mode.

Figure~\ref{fig:ann_frac} shows the fraction of the annihilation rate outside our benchmark neutron star as a function of inelastic mass splitting. We show this fraction for both $s$-wave and $p$-wave channels. For mass splittings $\delta \gtrsim \delta_{\rm min}$, the fraction grows until is maximal at $\delta \simeq \delta_{\rm max}$. For the velocity-dependent $p$-wave channel, the dark matter annihilates more efficiently when closer to the neutron star surface, where a higher velocity is attained. This results in a smaller fraction annihilating outside compared to the $s$-wave mode for a given splitting. Note that although the fraction annihilating outside grows with mass splitting, the overall signal at large mass splittings will be weaker because the capture efficiency is suppressed for splittings sufficiently close to the maximum value. 
\section{Neutrino and Gamma-Ray Signals}
\label{sec:nuandgammasignals}
If dark matter capture-annihilation equilibrium is attained, the annihilation rate around a neutron star is determined by the capture rate itself. From this we can construct the signal by integrating the capture rate from the neutron star population within the Galactic Center. To compare results across various possible combinations of neutron star and dark matter density profiles, it is useful to introduce the capture rate density as 
\begin{equation}
\rho_{C_\chi}(R) = \eta_{\rm NS}(R) \times C_\chi(R)~,
\end{equation}
where $\eta_{\rm NS}$ is the neutron star number density. Above, we have explicitly indicated that the capture rate $C_\chi$ for a neutron star is a function of Galactocentric distance $R$, as it is proportional to the dark matter density where the neutron star is positioned.

\begin{figure}[t]
    \centering
\includegraphics[width=1.0\linewidth]{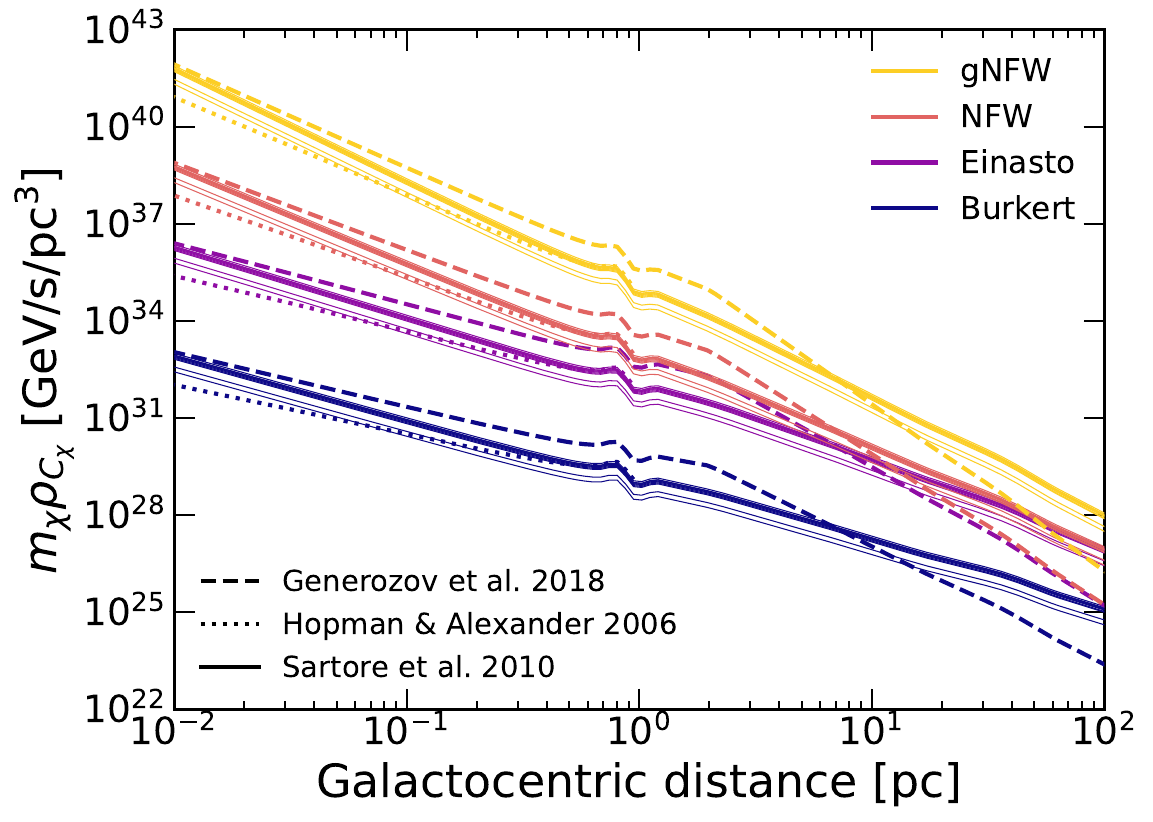}
    \caption{Capture rate density for the various combinations of neutron star distributions and dark matter density profiles as specified. Different solid lines for each dark matter profile correspond to the multiple fitting models discussed in Ref.~\cite{SartNS}. The power-law distribution of Ref.~\cite{Hopman:2006xn} is terminated beyond $\sim 1 \ \rm pc$.} 
    \label{fig:capture_profile}
\end{figure}

Figure~\ref{fig:capture_profile} shows the capture rate density for each combination of neutron star and dark matter distribution we consider. Overall, the cuspier gNFW profile predicts the largest dark matter capture rate on neutron stars at the Galactic Center, for any assumed neutron star distribution model. Degeneracies between the chosen dark matter and neutron star distributions only show up at Galactocentric distances $\gtrsim 10 \ \rm pc$, where the dark matter density is so small that neutron stars in this range do not significantly contribute to the total annihilation signal. As we show below, this means the strongest limits and projections are obtained for the gNFW profile. We also note that, for a fixed dark matter distribution, the variation between the different neutron star models is about a factor 5 at small Galactocentric distances. 

For any combination of neutron star and dark matter distribution, the capture rate density drops steeply away from the Galactic Center. Because of this, we find the dominant component of the signal comes from the innermost region of the Galactic Center. Therefore, for the purposes of computing fluxes of high energy annihilation byproducts, we model the neutron star population as a single point source, with a total annihilation rate 
\begin{equation}
   \Gamma_{\rm ann}^{\rm tot} = \frac{1}{2}\, \int_{R_{\rm min}}^{R_{\rm max}}\rho_{C_\chi}\!(R) \, 4\pi R^2 \, dR~.
\end{equation}
We take as a lower limit of integration $R_{\rm min} = 10^{-2} \ \rm pc$, in order to avoid the large uncertainty in dark matter and stellar velocities closer to the Galactic Center. The upper cutoff depends on the range of the neutron star distribution chosen. The maximum distance we integrate over is $R_{\rm max} = 50 \ \rm pc$, and we do not extrapolate any neutron star distribution beyond this point. In any case, most of the contribution to the signal is produced by the neutron stars closer to $R_{\rm min}$, where the dark matter density is considerably higher. 

Using this single point source approximation, the spectral flux of neutrinos or gamma-rays at Earth is 
\begin{equation}
    \frac{d\Phi_{j}}{dE_j}\bigg|_{\mathrm{ch}} = f_{\rm out} \times \frac{\Gamma^{\rm tot}_{\rm ann}}{4\pi D^2} \times \frac{dN_{{j}}}{dE_{j}} \bigg|_{\rm ch}~,
\label{eq:nu_gamma_flux}
\end{equation}
where $j = \nu_\alpha$ or $\gamma$ denotes the final observed state of a channel $ch$, $E_j$ is the energy, and $D \simeq 8.5 \ \rm kpc$ is the approximate Earth-Galactic Center distance. The factor $dN_{j}/dE_j |_{\rm ch}$ denotes the neutrino/gamma-ray spectrum per annihilation for a channel $\rm ch$. For sufficiently energetic gamma-rays, Eq.~\eqref{eq:nu_gamma_flux} is also multiplied by a survival probability that accounts for the attenuation through the interstellar medium (see below).

To span a range of spectral fluxes from soft to hard, we will consider three benchmark channels where dark matter annihilates exclusively to $b\bar{b}$, $\tau^+ \tau^-$ and $W^+ W^-$. We will also consider direct annihilation to neutrinos and gamma-rays to illustrate the maximum reach of our search using neutrino and gamma-ray telescopes, respectively. For the direct neutrino channel, we assume an equal contribution from each flavor. The neutrino and gamma-ray spectra are computed using {\texttt{HDMSpectrum}}~\cite{Bauer:2020jay}, which includes a state-of-the-art treatment of electroweak interactions. Note that, due to electroweak showers, a flux at energies below the dark matter mass is expected even for the direct neutrino/gamma-ray channel, which makes it possible to probe dark matter with mass larger than the maximum energy reach of an experiment in this case. At the same time, due to the electroweak interactions, we also expect a neutrino (gamma-ray) flux from the direct gamma-ray (neutrino) annihilation channel.

\subsection{Sensitivity of Neutrino Telescopes}
\label{sec:neutrino_signal}
We compute the sensitivity of IceCube and IceCube-Gen2~\cite{IceCube-Gen2:2020qha}, the next generation of IceCube at the South Pole. Detectors in the Southern Hemisphere like IceCube have a superior sensitivity to the Galactic Center at energies below hundreds of TeV, as the Earth shields the atmospheric muons, while a good view of the Southern Sky is also expected at higher energies due to the absence of neutrino flux attenuation by the Earth. An event selection also imposes a higher energy threshold, which can further suppress the atmospheric muon background. 
On the other hand, detectors in the Northern Hemisphere can bring complementary features, so we analyze these as well. There are various water Cherenkov telescopes either under construction or proposed in the Northern Hemisphere, such as Baikal-GVD, KM3NeT, P-ONE, TRIDENT, and the newly proposed HUNT~\cite{Baikal-GVD:2018isr,KM3Net:2016zxf,P-ONE:2020ljt,Ye:2022vbk,Huang:2023mzt}. 
Thus, we also report on the detectability of our signal using KM3NeT and TRIDENT. Baikal-GVD and P-ONE are expected to reach comparable sensitivity as KM3NeT with the current information of the detector installation and performance. In-ice/water Cherenkov detectors run out of sensitivity beyond $\sim 10 \ \rm{PeV}$ energies. For the detection of ultra-high-energy neutrinos above this threshold, we forecast the sensitivity of the proposed IceCube-Gen2 radio~\cite{IceCube-Gen2:2020qha}, which is expected to have the optimal sensitivity towards the Southern Sky.

Since we are considering a point source signal, we study track events induced by charged-current interactions of muon neutrinos, which have ideal pointing power. The expected signal track events in an energy bin $i$ can be written as 
\begin{equation}
    N_{\mu,i} = T \times \frac{1}{3}\sum_{\alpha = 1}^3\int_{E_{\nu_\mu,i}^{\rm min}}^{E_{\nu_\mu,i}^{\rm max}} A_{\rm eff}(E_{\nu_\mu}) \, \frac{d\Phi_{\nu_\alpha}}{dE_{\nu_\alpha}}\bigg|_{\mathrm{ch}} \! dE_{\nu_\mu}~,
\label{eq:event_number_nu_mu}
\end{equation}
where $A_{\rm eff}$ is the effective area which depends on the neutrino energy $E_{\nu_\mu}$ for the signal from a specific direction and $T$ is the exposure time which we set to 10 yr for our sensitivity estimation. The integration limits ${E_{\nu_\mu,i}^{\rm min}}$ and ${E_{\nu_\mu,i}^{\rm max}}$ correspond to the energy bounds of bin $i$. As the neutrino flux at detection is the flux after neutrino oscillation, we consider an equal contribution of each flavor in the total neutrino flux at Earth. When considering the high-energy neutrino telescopes, we use $A_{\rm eff}$ from the IceCube point-source data release~\cite{IceCube:2021xar}. The $A_{\rm eff}$ for KM3Net and IceCube-Gen2 is from the \texttt{PLE$\nu$M} framework which computes effective areas based on a detector location transfer and size scaling of the $A_{\rm eff}$ of IceCube~\cite{Schumacher:2021hhm}. The $A_{\rm eff}$ of TRIDENT is obtained from Figure~15 of Ref.~\cite{Ye:2022vbk}. For the ultra-high-energy regime, the diffuse neutrino flux sensitivity is computed with an expectation of 2.44 events per energy decade. Thus, we obtain the $A_{\rm eff}$ for IceCube-Gen2 radio from its reported energy-dependent sensitivity $\phi_{\rm{sens}}$ in Figure~19 of Ref.~\cite{IceCube-Gen2:2020qha}, by taking~\cite{Chianese:2021htv}
\begin{equation}
   A_{\rm {eff}} \! \left(E_\nu\right)=\frac{2.44 \, E_\nu}{4\pi T \, {\rm{ln}}(10) \, \phi_{\rm{sens}} \! \left (E_\nu \right)}~, 
\end{equation}
where the exposure time here has $T = 10 \ \rm yr$. 

We compute the sensitivity based on a likelihood ratio test 
\begin{align}   
\mathcal{L}  = e^{-N_{\mu}} \prod_i \frac{N^{N^{\rm obs}_{\mu,\,i}}_{\mu,\,i}}{N^{\rm obs}_{\mu,\,i}!}~,
 \label{eq:likelihood}
\end{align}
where $N_\mu = N_{\mu}^{\rm sig} + N_{\mu}^{\rm bkg}$ is the expected total number of $\mu$-tracks from the signal and the background (the latter is detailed below). 
The superscript $obs$ is used for the observed number of events and $i$ indicates the bin. By Wilks' theorem, we have $\chi^2 = -2 \, \mathrm{log}\left[\mathcal{L}\left(N_{\mu}^{\rm sig}=0\right)/\mathcal{L}\left(\hat{N}_{\mu}^{\rm sig}\right) \right]$, where $\hat{N}_{\mu}^{\rm sig}$ is the value that maximizes the ratio. Considering the $\chi^2$ distribution with 1 degree of freedom, we define our 90\% sensitivity as $\chi^2 = 2.7$. 

\begin{figure*}
    \centering
    \includegraphics[width=\textwidth]{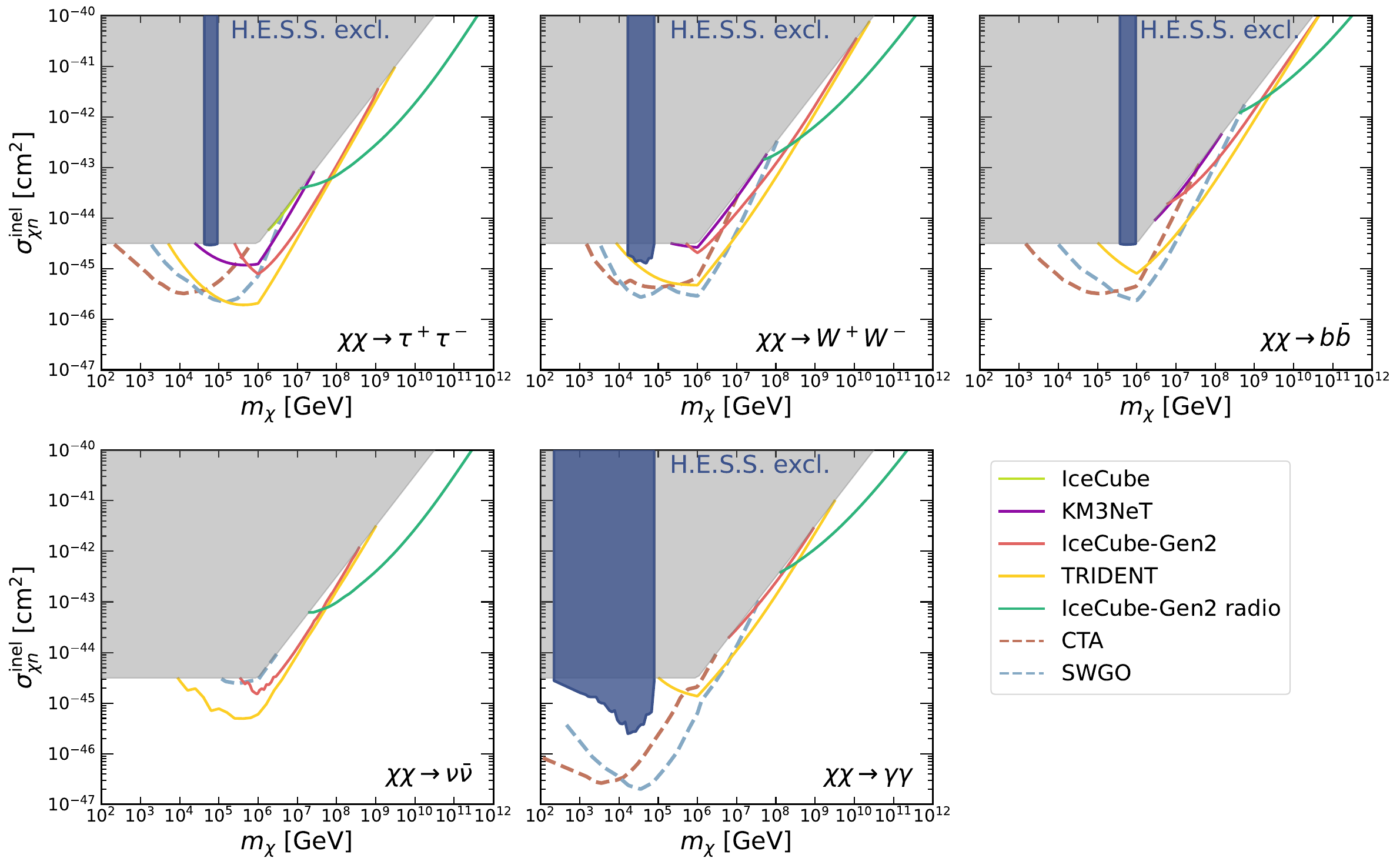}
    \caption{Inelastic dark matter-nucleon cross-section constraints derived from H.E.S.S.~Galactic Center gamma-ray observations \cite{HESS:2018pbp} (\textbf{blue}), assuming a gNFW profile and a fraction $f_{\rm out} = 0.95$ of dark matter annihilating outside neutron stars.  Also shown are 90\% sensitivities for neutrino telescopes (\textbf{solid}) with a 10~yr exposure in all cases, and projected 5$\sigma$-sensitivities of future Southern Hemisphere gamma-ray observatories (\textbf{dashed}) with a 50~hr exposure for CTA and 5~yr exposure for SWGO. Projections are drawn using the most optimistic neutron star density model. For reference, we indicate the inelastic cross-section for which approximately all the incoming dark matter is captured (\textbf{grey}); projections and bounds terminate on this line when their observational sensitivity threshold requires more dark matter annihilation than can be produced by all dark matter falling onto the Galactic Center neutron stars.}
    \label{fig:sensitivity_gnwf}
\end{figure*}

For high-energy neutrino detection, the main background is composed of atmospheric neutrinos and the diffuse astrophysical neutrino flux. We fix the atmospheric neutrino flux to the conventional flux given by the {\texttt{MCEq}} simulation with cosmic-ray flux model H3a, hadronic model SIBYLL-2.3c and atmosphere model NRLMSISE-00~\cite{MCEq,Riehn:2017mfm,picone2002nrlmsise}. 
For the diffuse astrophysical neutrino flux, we incorporate the best-fitted spectrum of astrophysical muon neutrinos~\cite{IceCube:2021uhz}. The angular resolution plays an important role in a point-source search for the background suppression power. The angular resolution of an in-water detector is expected to be better than an in-ice experiment due to less Cherenkov light scatterings. For our work, we consider the energy dependent angular resolutions from Ref.~\cite{IceCube:2019cia} for IceCube and IceCube-Gen2, and from Ref.~\cite{KM3Net:2016zxf} for other in-water experiments. In the ultra-high-energy range, the atmospheric background is negligible, but there is a diffuse cosmogenic neutrino flux expected from cosmic rays interacting with the cosmic microwave background and the extra-galactic background light~\cite{Berezinsky:1969erk}. In this case, we construct the background taking into account the predicted cosmogenic neutrino flux from cosmic-ray studies~\cite{Heinze:2019jou}. The direction of a neutrino event in a future Askaryan detector can reach within $1^\circ$~\cite{Allison:2017jpy}, and so we take $1^\circ$ for our work when estimating the sensitivity of IceCube-Gen2 radio. Since we consider neutrons stars up to 50~pc from the Galactic Center, which corresponds to $\sim 0.3^\circ$, we also include this angular extension when computing the background for both high-energy and ultra-high-energy detection. 

\subsection{Sensitivity of Gamma-Ray Observatories}
\label{sec:gamma_signal}
There are various running or proposed ground-based gamma-ray experiments with sensitivity to very-high-energy gamma-rays. We will focus on those located in the Southern Hemisphere, which have exposure to the Galactic Center. Note that at energies above tens of TeV, attenuation due to the pair production mainly with the cosmic microwave background and the infrared emission by dust become non-negligible for gamma-rays as they propagate through the interstellar medium~\cite{Fang:2021ylv}. We incorporate this correction by multiplying Eq.~\eqref{eq:nu_gamma_flux} by a survival probability given by
\begin{equation}
    P_{\rm surv} = \mathrm{exp}\left(-\tau_{\gamma\gamma} \! \left( E_\gamma\right)\right)~,
\end{equation}
where $\tau_{\gamma\gamma}(E_\gamma)$ is the energy-dependent optical depth between the Galactic Center and an observer at the Milky Way. We use the optical depth data from Ref.~\cite{Fang:2021ylv}.

We compute inelastic cross-section constraints from existing observations performed by the High Energy Stereoscopic System (H.E.S.S.), which currently has the leading sensitivity to very-high-energy gamma-rays toward the Southern Sky. To draw these limits, we compare our predicted fluxes to the energy differential fluxes reported in the Galactic Plane Survey~\cite{HESS:2018pbp}, which roughly cover an energy range of about $0.1 - 100 \ \rm TeV$, and require that our flux must not exceed the observed flux in any energy bin. 

We also forecast the sensitivity of the Southern Array of the Cherenkov Telescope Array (CTA) and the Southern Wide-field Gamma-ray Observatory (SWGO), which will significantly improve observations of the Galactic Center in the near future. The energy differential point-source sensitivities of these telescopes have been evaluated in Refs.~\cite{Maier:2019afm, Albert:2019afb}, which correspond to a total observation time of 50 hours with the Southern Array of CTA and a 5~yr exposure for SWGO. The projected sensitivity for each case is then obtained following the same procedure as with H.E.S.S. observations. Note that the reported sensitivities in Refs.~\cite{Maier:2019afm, Albert:2019afb} correspond to a 5$\sigma$ detection while constraints on dark matter studies are usually at the 90\% confidence level. 

\subsection{Results}
Figure~\ref{fig:sensitivity_gnwf} shows our computed dark matter-nucleon inelastic cross-section limits based on H.E.S.S. gamma-ray data and projected sensitivities of upcoming neutrino and gamma-ray observatories, for the gNFW profile. These are shown for the annihilation channels specified at the start of Sec.~\ref{sec:nuandgammasignals}. For concreteness, we have assumed all neutron stars to be given by our benchmark (see Sec.~\ref{sec:gc_mod}), and fixed $f_\mathrm{out}=0.95$ (all projected limits scale linearly with this parameter). This corresponds to an inelastic mass splitting of about $190 \ \rm MeV$ ($239 \ \rm MeV$) for $s$-wave ($p$-wave) annihilation mode, $cf.$ Figure~\ref{fig:ann_frac}. For reference, we also show the cross-section at which nearly all the incoming dark matter is captured by a neutron star. Sensitivity is lost above this line because the capture rate of neutron stars is maximized, and thus higher cross-sections do not physically increase the annihilation signal. In all cases, we have used the most optimistic neutron star distribution. However we note that the variation of our results between all neutron star distributions is about a factor $\sim 5$. Advances in our understanding of star formation in the Milky Way's nuclear star cluster will reduce this uncertainty.

\begin{figure}[t!]
    \centering
    \includegraphics[width=1.0\linewidth]{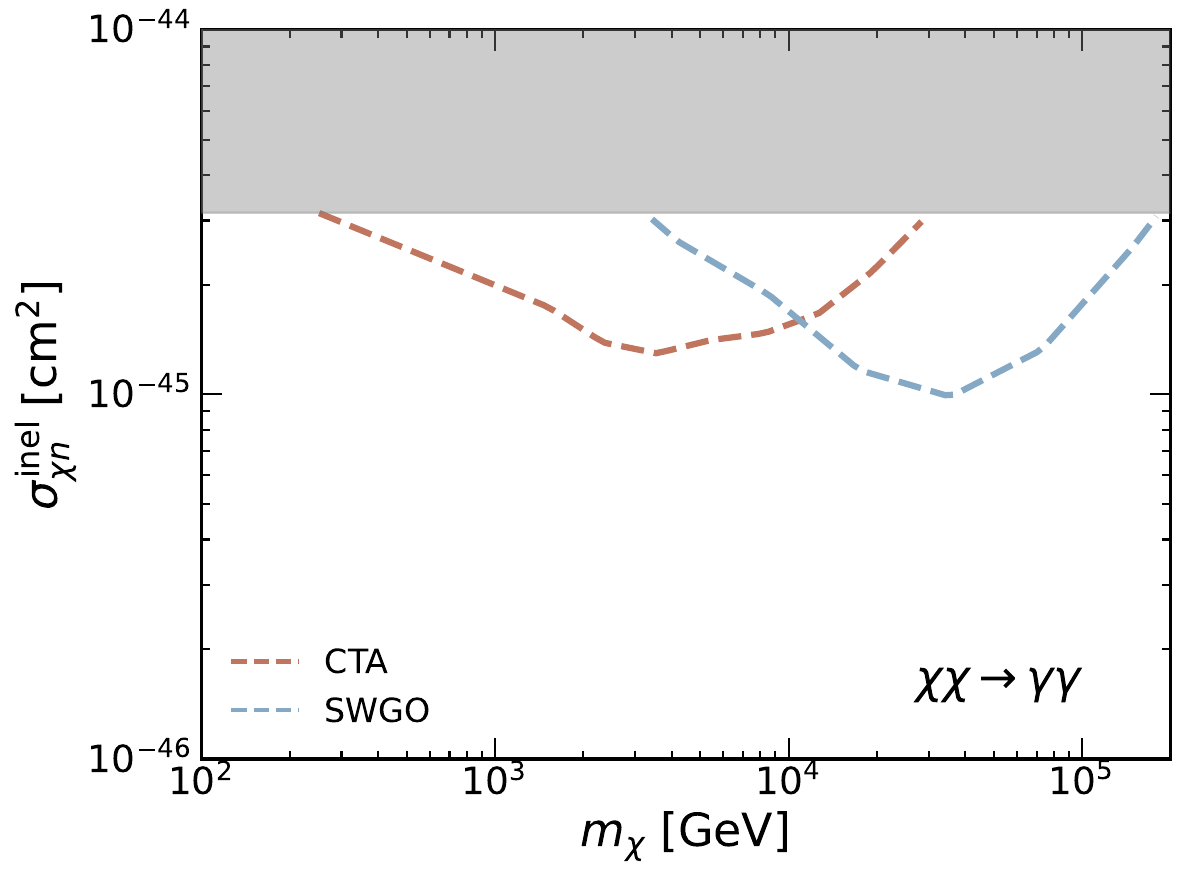}
    \caption{Projected sensitivities for CTA and SWGO to inelastic dark matter-nucleon cross-sections shown for the direct gamma-ray annihilation channel, assuming an NFW profile. Otherwise same as Fig.~\ref{fig:sensitivity_gnwf}.}
    \label{fig:sensitivity_nfw_gamma}
\end{figure}

As expected, the strongest sensitivity is attained for the steeper gNFW profile, with H.E.S.S. limits reaching down to $\sim 3 \times 10^{-46} \ \rm cm^2$ while the strongest projections achieve $\sim 2 \times 10^{-47} \ \rm cm^2$ for direct annihilation to gamma-rays. Note that, for both the gNFW and NFW profiles, the halo annihilation signal can be comparable to the signal sourced by neutron stars for $s$-wave annihilation below $\sim \rm PeV$ masses. For velocity-dependent annihilation, such as $p$-wave and higher modes, the neutron star signal will be significantly stronger than its halo counterpart. Using the conservative assumptions laid out in this first exploratory analysis, we did not find robust search prospects for an NFW profile, except for some limited sensitivity in the mass range $\sim 1 - 100 \ \rm TeV$, using the future reach of CTA and SWGO in the direct gamma-ray channel, as shown in Figure \ref{fig:sensitivity_nfw_gamma}. It is possible that future analyses treating the thermalization of inelastic dark matter and capture-annihilation equilibration using less conservative assumptions could find gamma-ray and neutrino flux sensitivity for an NFW profile. For the Einasto and Burkert profiles, the dark matter content in the Galactic Center is too low to achieve sensitivity in any channel. 

The sharp loss in constraining power for the direct gamma-ray channel above $\sim 100 \ \rm TeV$ is due to the observable signal switching from a line-emission-like signal at the dark matter mass to an electroweak shower at lower energies. This also applies to the $W^+W^-$ channel. Neutrino experiments, on the other hand, have ample sensitivity to energies beyond the PeV scale, and thus achieve the best projections for heavy inelastic dark matter. For the direct neutrino channel, our forecasted sensitivity reaches $\sim 4 \times 10^{-46} \ \rm cm^2$. Depending on the dark matter profile, IceCube-Gen2 radio will obtain leading constraints on the inelastic cross-section for ultra-heavy dark matter. The Southern Array of CTA and SWGO will also achieve a sensitivity comparable to that of neutrino experiments up to $\sim 10 \ \rm PeV$ masses, except for the direct neutrino channel to which neutrino experiments are naturally more sensitive. 

\section{Models for Inelastic Dark Matter}
\label{sec:dpmodel}
Inelastic dark matter models have been widely considered in the past, for example, as a way of addressing outstanding astrophysical excesses \cite{Batell:2009vb,Finkbeiner:2007kk,Finkbeiner:2009mi,Pospelov:2007xh} or small-scale structure problems \cite{Zhang:2016dck,Blennow:2016gde,Alvarez:2019nwt} while remaining consistent with null results from direct detection searches. In the most minimal realization, a small Majorana mass term is introduced, which splits the Dirac dark matter into two neutral components with a mass difference set by the Majorana mass. Motivated by the self-interacting dark matter scenario, we illustrate the applicability of our search with a simple inelastic dark photon-mediated dark matter model \cite{Bramante:2016rdh}
\begin{align}
    \label{eq:dp_lag}
    \mathcal{L} & = \mathcal{L}_{\rm SM} + |D_{\mu}\phi|^2 + V(\phi) - \frac{1}{4} V_{\mu \nu}^2 + \kappa V_{\mu}\partial_{\nu}F^{\mu \nu} \\ & - \frac{1}{2} m_V^2 V_\mu^2  + \bar{\chi}\left(i D_{\mu}\gamma_{\mu}-m_{\chi}\right)\chi + \left(\lambda_D \phi \chi^T C^{-1}\chi + {\rm h.c.} \right)~, \nonumber
\end{align}
where $V_\mu$ ($V_{\mu\nu}$) is the new $U(1)_D$ gauge boson (field strength tensor) which mixes with the SM photon with kinetic mixing parameter $\kappa$, $D_{\mu} \equiv \partial_{\mu} + i e_D V_{\mu}$ where $e_D$ is the charge of the dark matter under $U(1)_D$, and $C$ is the charge conjugation matrix. The mass splitting is achieved through spontaneous symmetry breaking of the $U(1)_D$ by a scalar $\phi$ that also couples to the dark matter. We focus on the heavy dark photon parameter space of masses $m_V \gtrsim 1 \ \rm GeV$, which currently is relatively unconstrained. The value of $\alpha_D = e_D^2/4\pi$ can be fixed by requiring for example standard thermal freeze-out. However, to illustrate our point we remain agnostic about the cosmological formation history.  

We now show how all the necessary conditions for our search to be applicable are generically fulfilled:
\begin{itemize}
\item For reasonable parameters, the resulting splitting between the mass eigenstates can be of order
    \begin{equation}
      \delta \simeq \lambda_D v_\phi = 100 \ {\rm MeV} \left(\frac{\lambda_D}{0.1}\right) \left(\frac{v_\phi}{\rm GeV}\right)~,
     \end{equation} 
where $v_\phi$ is the vacuum expectation value of the scalar. At this mass splitting, endothermic scattering in direct detection searches is kinematically forbidden for virialized halo dark matter, while in a typical neutron star it can be captured but will only partially thermalize with it. 

\item Through the dark photon portal, dark matter scatters off charged particles in a neutron star, which comprise about a few percent of the total particle number (see $e.g.$ \cite{Lattimer:2004pg}). The inelastic dark matter-proton cross-section dominates over the electron cross-section and is given by 
\begin{align}
   \ \ \ \ \ \ \ \sigma_{\chi p}^{\rm inel} & \ =  \frac{16 \pi \alpha \alpha_D \kappa^2 m_p^2}{m_V^4} \\
   \simeq & \ 1.2 \times 10^{-43}\ {\rm cm^2} \left(\frac{\kappa}{10^{-5}}\right)^2 \left(\frac{\alpha_D}{0.1}\right)\left(\frac{10 ~{\rm GeV}}{m_V}\right)^4 \nonumber
\end{align}
which can be well above the necessary value to maximize capture, even accounting for the said reduction in the number of scattering targets for this specific SM portal. By contrast, the loop-level elastic scattering cross-section is 
\begin{align}
\ \ \ \ \ \ \ \sigma_{\chi p}^{\rm elas} & \ =  \frac{\alpha^2 \alpha_D^2 \kappa^4 m_n^4 f_q^2}{\pi m_V^6}  \\
    \simeq & \ 5.1 \times 10^{-63} \ {\rm cm^2} \left(\frac{\kappa}{10^{-5}}\right)^4 \left(\frac{\alpha_D}{0.1}\right)^2  \left(\frac{10 \, {\rm GeV}}{m_V}\right)^6 \nonumber
\end{align}
where $f_q \sim 0.1$ is a hadronic matrix element \cite{Bramante:2016rdh}. These cross-sections imply efficient inelastic capture in neutron stars, but thermalization is extremely slow if the inelastic transition becomes kinematically suppressed. Note that for the dark photon mass range considered, the heavy state will be long-lived, as it is unable to decay by dark photon emission. Thus, in the multiscatter capture regime, it must exothermically decay back into the light state. Note that the kinetic mixing values we consider lie significantly below current limits from visible final state searches at accelerators \cite{BaBar:2014zli,LHCb:2019vmc,CMS:2019buh}.

\item The captured dark matter annihilates to dark photons, which then decay to charged particles and subsequently produce gamma-rays. Dark photons with masses $m_V \gg 1 \ \rm GeV$ decay to Standard Model leptons in a time of order $(\alpha \kappa^2 m_V/3)^{-1} \sim 10^{-13} \ {\rm s} \, (10^{-5}/\kappa)^2 \, ({10 \ \rm GeV}/m_V)$, and for hadronic final states this lifetime changes by an order one factor given by the hadron-to-muon production cross-section ratio in $e^{+}e^{-}$ annihilation, see $e.g.$ \cite{Fabbrichesi:2020wbt}. The resulting decay length will be much shorter than the neutron star size unless the boost factor is of order $m_\chi/m_V \gtrsim 10^{8}$, which is not achieved in a sizable region of the parameter space considered. This implies that only partially thermalized inelastic dark matter annihilating outside a neutron star will source a potentially observable signal. 
\end{itemize}

This simple model realization illustrates the wide class of models for which our search would have ample sensitivity, while direct detection rates would be severely suppressed. Although a neutron star kinetic heating search would also be applicable to this scenario, we emphasize that it is contingent upon the existence of nearby faint neutron stars and lengthy observation times with infrared telescopes. 

\section{Conclusions}
\label{sec:conc}
We have analyzed for the first time the annihilation of inelastic dark matter outside neutron stars. This effect arises because, after being captured, the dark matter particles are unable to fully thermalize with the neutron star due to inelastic kinematics. If elastic interactions are suppressed relative to the inelastic channel, these can remain in long-lived orbits that extend beyond the neutron star volume, allowing for a sizable fraction to annihilate outside. For interstate splittings approximately within $45 - 285 \ \rm MeV$, a significant fraction of the captured dark matter annihilates outside the volume of a typical neutron star, producing a potentially observable signal for neutron stars in dark matter-dense environments. We have detailed this process in a model-independent manner, and estimated the rate of annihilation proceeding outside neutron stars as a function of the interstate mass splitting.

We have investigated the detection prospects of this annihilation signal by targeting the neutron star population in the Galactic Center, where the dark matter content is robustly expected to be high and neutron star-focused annihilation can dominate over halo annihilation. Specifically, we have computed the resulting gamma-ray and neutrino signals produced from a variety of annihilation channels. Naturally, these predictions depend on the assumed dark matter and neutron star distribution profile. For a cuspy generalized NFW dark matter profile motivated by adiabatic contraction studies, and various neutron star distributions, we have placed constraints based on H.E.S.S.~Galactic Center observations on the inelastic dark matter-nucleon cross-section. These constraints can reach down to $\sim 3 \times 10^{-46} \ \rm cm^2$ in the case of direct annihilation to photons, for dark matter masses ranging $10^2 - 10^5 \ \rm GeV$. Our procedure for setting the H.E.S.S.~bound required the gamma-ray flux from dark matter annihilation exceed the total gamma-ray flux from the Galactic Center; future analyses may improve on this method and find a stronger bound. We have also computed the sensitivity of future gamma-ray and neutrino observatories to this signal. Depending on the annihilation channel, as well as the assumed dark matter and neutron star distributions, future neutrino and gamma-ray telescopes will reach inelastic cross-sections as low as about $\sim 2 \times 10^{-47} \ \rm cm^2$. 

This study demonstrates the use of neutron star populations in high dark matter density systems as a means of probing inelastic dark matter models, across a mass range that spans $\sim$10 orders of magnitude. In this scenario, nuclear scattering in direct detection experiments is either kinematically forbidden or has a minuscule rate, potentially making these objects the only viable way of probing these models. Notably, this search can complement neutron star heating searches, which have been previously explored but crucially depend on the proximity of old neutron stars, require the allocation of significant infrared telescope observation time, and might be masked by other internal neutron star heating mechanisms.  

\begin{acknowledgments}
We thank Rebecca Leane and Nick Rodd for helpful comments and discussions. We also thank Ke Fang for providing the gamma-ray optical depth table. JFA is supported in part by the U.S. Department of Energy under Contract DE-AC02-76SF00515. JB, QL and NT are supported by the Arthur B. McDonald Canadian Astroparticle Physics Research Institute, with equipment funded by the Canada Foundation for Innovation and the Province of Ontario, and housed at the Queen’s Centre for Advanced Computing. Research at Perimeter Institute is supported by the Government of Canada through the Department of Innovation, Science, and Economic Development, and by the Province of Ontario.
\end{acknowledgments}

\appendix

\section{Neutron Star Structure}
\label{app:ns_struc}
For a static and spherically-symmetric neutron star, the spacetime interval is expressed as
\begin{equation}
    ds^2 = g_{tt}(r) dt^2 + g_{rr}(r) dr^2 - r^2 d\Omega^2~,
\end{equation}
where $d\Omega^2 = d\theta^2 + \sin^2 \! \theta \, d\varphi^2$. To determine the neutron star density profile and the metric components in its interior, we numerically solve the TOV equation
\begin{equation}
    \frac{dP}{dr} = - \frac{G\rho m}{r^2} \left(1+\frac{P}{\rho}\right)\left(1+\frac{4\pi P r^3}{m}\right) \left(1-\frac{2Gm}{r}\right)^{-1}~,
    \label{eq:tov1}
\end{equation}
\begin{equation}
    \frac{dm}{dr} = 4\pi r^2 \rho~,
    \label{eq:tov2}
\end{equation}
\begin{equation}
    \frac{d\Phi}{dr} = - \frac{1}{\rho} \frac{dP}{dr} \left(1+\frac{P}{\rho}\right)^{-1}~.
    \label{eq:tov3}
\end{equation}
Above, $P(r)$ is the pressure, $\rho(r)$ is the energy density, $m(r)$ is the enclosed gravitational mass within $r$, and $g_{tt}(r) = \exp(2\Phi(r))$. The remaining interior metric component is $g_{rr}(r) = - \left(1 - 2Gm(r)/r \right)^{-1}$. Outside the neutron star, the solution is matched to the Schwarzchild metric, $i.e.$ $g_{tt}\left(r \geq R_{\rm NS}\right) = 1 - 2GM_{\rm NS}/r$ and $g_{rr}\left(r \geq R_{\rm NS}\right) = - g_{\rm tt}^{-1}(r)$. The above system is closed when an equation of state, $i.e.$ a constitutive relation of the form $P=P(\rho)$, and the central energy density are specified. It can then be solved through standard numerical methods. For our chosen BsK-21 equation of state, fixing the central energy density to $\rho_c \simeq 7.85 \times 10^{15} \ \rm g \ cm^{-3}$ yields our benchmark neutron star of mass $M_{\rm NS} \simeq 1.5 M_{\odot}$ and radius $R_{\rm NS} \simeq 12.55 \ \rm km$. 

\section{Dark Matter Orbital Period and Velocity}
\label{app:dm_orb}
We derive general relations used in the main text for the orbital period and the velocity of a dark matter particle. In what follows, we will assume the total energy of the particle remains fixed. We start from the Hamilton-Jacobi equation (see $e.g.$ \cite{landau2013classical})
\begin{equation}
    g^{\mu \nu}\frac{\partial S}{\partial x^{\mu}} \frac{\partial S}{\partial x^{\nu}} - m_\chi^2= 0~,
    \label{eq:HJeq}
\end{equation}
where the metric components $g^{\mu \nu} = g_{\mu \nu}^{-1}$ are determined from Eqs.~\eqref{eq:tov1}$-$\eqref{eq:tov3} above, and $S(x)$ is the action evaluated along the physical trajectory (for abbreviation, $x = (t,r,\theta,\varphi)$). This equation is solved by separation of conjugate variables associated with constants of motion. Using the fact that the orbits are contained within a plane, we fix $\theta = \pi/2$ without losing generality. The ansatz solution is expressed as 
\begin{equation}
    S(x) = \varepsilon \, t - M \, \varphi + S_r(r) + \rm const.~,
    \label{eq:HJeq_sol}
\end{equation}
where $\varepsilon$ and $M$ are the dark matter particle's total energy and orbital angular momentum, respectively. This ansatz is introduced in Eq.~\eqref{eq:HJeq}. Isolating the derivative $S_r(r)$, one can integrate the action function up to an arbitrary constant fixed by the initial condition,
\begin{equation}
    S_r(r)= \int \left[g_{rr}(r)\left(m_\chi^2 + \frac{M^2}{r^2} - \frac{\varepsilon^2}{g_{tt}(r)} \right)\right]^{1/2} \, dr~.
    \label{eq:HJeq_sol2}
\end{equation}
We numerically integrate the above using the metric components for our benchmark neutron star profile, $cf.$ App.~\ref{app:ns_struc}. 

From Eq.~\eqref{eq:HJeq_sol}, it is possible to compute the orbital period and velocity as a function of the dark matter energy, which are required inputs to analyze how thermalization proceeds after capture. For a given $\varepsilon$ and $M$, the orbital time between to coordinates $r$ and $r_0$ is given by \cite{landau2013classical}
\begin{equation}
    t - t_0= \frac{\partial S_r}{\partial\varepsilon}~,
\end{equation}
where $S_r(r)$ is integrated from $r_0$ to $r$. We compute the orbital period as a function of energy from this relation. To compute the velocity measured by a local observer, we use the relation between energy and the time component of the 4-velocity,
\begin{equation}
    \varepsilon = m_\chi \, g_{tt}(r) \, \frac{dt}{ds}~.
\end{equation}
For a constant gravitational field, we may write $ds = \sqrt{g_{tt}(r) dt^2 - dz^2}$, where $dz$ is the proper radial length measured by a local observer. Defining $v = dz/d\tau$, we have
\begin{equation}
    \varepsilon = m_\chi \left(\frac{g_{tt}(r)}{1-v^2}\right)^{1/2}~.
    \label{eq:app_energ}
\end{equation}
Since we are assuming here that the energy $\varepsilon$ is constant once the inelastic transition becomes suppressed, we have $(\varepsilon/m_\chi)^2 = {\rm const.} = g_{tt}(r) / (1-v^2)$. At a turning point $r = r_f$, the velocity vanishes and thus we have the relation
\begin{equation}
    g_{tt}(r_f) = \left(\frac{\varepsilon}{m_\chi}\right)^2
\end{equation}
between the turning point and the energy. Using this relation in Eq.~\eqref{eq:app_energ} and solving for $v$, we have 
\begin{equation}
    v(r) = \left(1-\frac{g_{tt}(r)}{g_{tt}(r_f)}\right)^{1/2}~.
\end{equation}
As a consistency check, note that in the limit $r_f \rightarrow \infty$, setting $r = R_{\rm NS}$ yields $v(R_{\rm NS}) = \sqrt{2 G M_{\rm NS}/R_{\rm NS}}$, which corresponds to the usual escape velocity expression used throughout the literature. \\
\section{Inelastic Dark Matter Profile}
\label{app:inel_prof}
We now estimate the resulting inelastic dark matter profile around a neutron star using simple kinetic theory arguments. In what follows, we will assume no angular momentum for simplicity. Furthermore, we will only consider the dark matter population that has lost enough energy so that the inelastic transition becomes kinematically forbidden. These are conservative choices because particles with non-zero angular momentum will generally spend a greater time orbiting outside the object, and we are also neglecting a small but finite population of dark matter particles that are at any given time in an earlier stage of thermalization. So, our results underestimate the true dark matter density outside of the neutron star that contributes to the overall annihilation signal.   

When the dark matter particles have lost enough energy, we assume elastic interactions are sufficiently suppressed that we may approximate the system as collisionless. In this regime, all the dark matter particles will be in approximately steady-state orbits, with randomly distributed phases and the same energy determined by the inelastic mass splitting. The occupation number of a phase space volume is invariant for any local Lorentz observer along the worldline of the dark matter particles \cite{landau2013classical}. Thus, a given dark matter particle in its rest frame sees a constant number density associated with the total number of particles with the same 3-momentum. Upon boosting to the frame of a local observer in the neutron star, that number density receives a factor $\sqrt{1-v(r)^2}$, $cf.$ Eq.~\eqref{eq:v_loc_main}. We then parameterize the number density as
\begin{equation}
    n_\chi(r) = n_\chi^{0} \left(1 - v(r)^2\right)^{1/2}
    \label{eq:app_DMdens}
\end{equation}
where $n_{\chi}^{0}$ is a normalization constant which can be fixed through the condition
\begin{equation}
    \int_0^{r_f} 4\pi r^2 n_{\chi}(r) dr =
    \begin{cases}
      C_{\chi} t_{\rm NS}~, \ \ t_{\rm NS} \ll t_{\rm eq} \\
      \\
      N_{\rm eq}~, \ \ t_{\rm NS} \gg t_{\rm eq}\\
    \end{cases}
\end{equation}
Above, $t_{\rm NS}$ is the lifetime of the neutron star, which we assume to be order gigayears, and $t_{\rm eq}$ is the timescale on which capture-annihilation equilibrium is reached, $cf.$ Eq.~\eqref{eq:N_eq_main}. \\
\bibliographystyle{JHEP}
\bibliography{apssamp}
\end{document}